\documentclass[journal]{IEEEtran}

\usepackage{graphicx}
\usepackage{subcaption}
\ifCLASSINFOpdf
\else
\fi
\usepackage{amsmath}
\usepackage{amsfonts}
\usepackage{color}

\usepackage[linesnumbered,ruled,vlined]{algorithm2e}

\usepackage{amsthm}

\theoremstyle{remark}

\begin{document}
%
\title{Fast Convolutional Dictionary Learning off the Grid}
%
%
%

\author{Andrew~H.~Song,~\IEEEmembership{Student Member,~IEEE,}
        Francisco~J.~Flores,
        and~Demba~Ba,~\IEEEmembership{Member,~IEEE}
\thanks{A. H. Song is with Electrical Engineering and Computer Sciences, Massachusetts Institute of Technology (MIT), Cambridge, MA (e-mail: andrew90@mit.edu)}
\thanks{F. J. Flores is with Massachusetts General Hospital (MGH) and Harvard Medical School, Boston, MA}
\thanks{D. Ba is with School of Engineering and Applied Sciences, Harvard University, Cambridge, MA (e-mail: demba@seas.harvard.edu)}
\thanks{Manuscript submitted July, 2019.}}

\markboth{}%
{Shell \MakeLowercase{\textit{et al.}}: Bare Demo of IEEEtran.cls for IEEE Journals}
%



\maketitle

\begin{abstract}
Given a continuous-time signal that can be modeled as the superposition of localized, time-shifted events from multiple sources, the goal of Convolutional Dictionary Learning (CDL) is to identify the location of the events--by Convolutional Sparse Coding (CSC)--and learn the template for each source--by Convolutional Dictionary Update (CDU). In practice, because we observe samples of the continuous-time signal on a uniformly-sampled grid in discrete time, classical CSC  methods can only produce estimates of the times when the events occur on this grid, which degrades the performance of the CDU. We introduce a CDL framework that significantly reduces the errors arising from performing the estimation in discrete time. Specifically, we construct an expanded dictionary that comprises, not only discrete-time shifts of the templates, but also interpolated variants, obtained by bandlimited interpolation, that account for continuous-time shifts. For CSC, we develop a novel computationally efficient CSC algorithm, termed Convolutional Orthogonal Matching Pursuit with interpolated dictionary (COMP-INTERP). We benchmarked COMP-INTERP to Contiunuous Basis Pursuit (CBP),  the state-of-the-art CSC algorithm for estimating off-the-grid events, and demonstrate, on simulated data, that 1) COMP-INTERP achieves a similar level of accuracy, and 2) is two orders of magnitude faster. For CDU,  we derive a novel procedure to update the templates given sparse codes that can occur both on and off the discrete-time grid. We also show that 3) dictionary update with the overcomplete dictionary yields more accurate templates. Finally, we apply the algorithms to the spike sorting problem on electrophysiology recording and show their competitive performance.
\end{abstract}

\begin{IEEEkeywords}
Convolutional Dictionary Learning, Convolutional Sparse Coding, Convolutional Orthogonal Matching Pursuit, Non-integer shift
\end{IEEEkeywords}

%
\IEEEpeerreviewmaketitle

\section{Introduction}
%
%
%
%

\IEEEPARstart{I}{n} recent years, the problem of decomposing an observed signal into a sparse linear combination of elements drawn from a known dictionary, often referred to as sparse approximation \cite{SparseApprox}, has been of great interest to the signal processing community. Specifically, representing the signal as the superposition of time-shifted (or shift-invariant) templates with local support has received special attention \cite{Lewicki1999}. This is due to the observation that many examples of real-world signals can be modeled in this manner. For instance, signals arising from electrophysiological recordings of neural activity can be modeled as the sum of distinct action potentials produced by the neurons near recording electrodes \cite{Sahani1998}. In studies involving electroencephalography (EEG) or magnetoencephalography (MEG), there is growing evidence that the signal should be studied in terms of the aggregation of transient events with specific templates \cite{Cole2017}.        

A generative model for these signals is the convolution in continuous time between templates, a collection of which is referred to as \textit{dictionary}, and a set of scaled and time-shifted delta functions, referred to as \textit{codes}. The amplitude and the location of each code correspond to the magnitude and the time when an event occurs, respectively. Given an observed signal, the goal of \textit{Convolutional Dictionary Learning} (CDL) frameworks is to estimate the templates and the codes under the generative model, and with sparsity constraints on the codes. These frameworks typically alternate between two steps, a \textit{Convolutional Sparse Coding} (CSC) step to estimate the codes, and a \textit{Convolutional Dictionary Update} (CDU) step to estimate the dictionary.

One of the drawbacks of existing CDL frameworks is the assumption that the signal of interest occurs in discrete time, when in fact the underlying signal occurs in continuous time. The discrete-time approximation of the generative model introduces errors known as time-quantization errors \cite{McGill1984}. Specifically, if an event in the continuous-time model were to occur at a time that does not coincide with any point on the discrete-time sampling grid, the CSC step would inaccurately identify the event as occurring at a time on the grid. Increasing the sampling rate to obtain a finer grid is a viable solution, but is not always possible for two reasons. First, practically, the sampling rate is limited by hardware specifications and the amount of memory available for storage. Second, the large number of samples will increase the computation time of the CSC and CDU steps.

Continuous Basis Pursuit (CBP) \cite{CBP}, a convex sparse regression framework, was introduced to address such limitations of the conventional approaches for CSC. The CSC step of CBP begins with the derivation, from the original dictionary, of an alternate dictionary that approximates the subspace of continuous-time-shifted copies of the templates (the continuous-time shifts are not necessarily multiples of the sampling interval). Then, using the new dictionary, it solves a $\ell_1$-regularized convex regression problem. Despite the method being more accurate, compared to discrete-time approaches, in estimating the times when events occur, it does not scale well with the size of modern datasets. Moreover, because CBP does not include a CDU step, it is not a CDL framework. 

With the increasing ability to record larger datasets, there is a need for an efficient CDL framework that addresses the time-quantization issue. As our primary motivation is efficiency, we focus on extending CDL frameworks based on greedy methods, which are known to be less computationally demanding than basis pursuit and $\ell_1$-regularized methods \cite{TROP2007}, \cite{Needell2009}, \cite{Kunis2008}. To address the time-quantization issue, we build an overcomplete and interpolated dictionary that accounts for events that occur off the discrete-time sampling grid. Unlike in CBP, this approach also allows us to learn the templates while incorporating the amplitudes and the times of events that occur off the grid.

Our contributions are the following:
\begin{itemize}
	\item \textbf{A fast convolutional greedy pursuit algorithm} We introduce an efficient algorithm for convolutional greedy pursuit under a discrete-time generative model. When the events from the continuous-time model occur on the sampling grid, we term this algorithm Convolutional Orthogonal Matching Pursuit (COMP). COMP is much faster than the alternative based on the basis-pursuit algorithm. Compared to Convolutional Matching Pursuit (CMP), the computational overhead from COMP is negligible (Table \ref{table:computation1}, \ref{table:computation2}).
	
	\item \textbf{A CSC framework that handles non-integer delays} We introduce a discrete-time generative model that accounts for events from the continuous-time model that do not occur on the discrete-time sampling grid. This model is inspired by the concept of bandlimited interpolation in digital signal processing. To perform the CSC, we extend COMP and call the resulting algorithm COMP-INTERP. Compared to conventional CSC frameworks, COMP-INTERP is more accurate in identifying the times when of off-the-grid events occur and achieves an accuracy similar to CBP (Figure \ref{fig:csc_comparison}). Moreover, COMP-INTERP is much faster than CBP (Table \ref{table:computation1}, \ref{table:computation2}).
	
	\item \textbf{A dictionary update framework that handles non-integer delays} For the first time, we introduce a CDU algorithm that accounts for estimates of the sparse codes from the CSC step that correspond to events off the grid. When compared to conventional dictionary update algorithms, the templates learned from our approach are more accurate (Figure \ref{fig:cdl}). In addition, our algorithm is more robust to varying noise levels (Figure \ref{fig:cdl_snr}).
	
\end{itemize}

The rest of our treatment begins in Section \ref{section:generative}, where we introduce the generative model of interest and formulate the CDL objective functions. In Section \ref{section:background}, we review existing work relevant to our CDL framework. In Section \ref{section:comp}, we introduce an efficient framework, termed COMP-INTERP, for performing CSC while accounting for events that occur off the grid. In Section \ref{section:cdl}, we introduce a CDU step that can handle events off the grid. We used simulated and real datasets to demonstrate the performance of our algorithms in Section \ref{section:experiments}. We conclude in Section \ref{section:conclusion}.

\section{Generative Model and Problem setup}\label{section:generative}
\label{gen_inst}

\subsection{Notations}
Table \ref{table:notations} summarizes our notational conventions. We introduce additional notation as necessary, at the beginning of the section that uses it first. We use the expressions \textit{event off the grid} and \textit{event with a non-integer delay} interchangeably.

\begin{table}[!ht]
	\caption{Notations}
	\centering
	\begin{tabular}{c l}
		Symbol & Description\\
		\hline\hline
		\\[-0.6em]
		$\mathbf{H}$ & Matrix\\
		$\mathbf{h}$ & Vector\\
		$\mathcal{S}$ & Set\\
		$\mathbf{H}_i$ & $i^{\text{th}}$ column from $\mathbf{H}$\\
		$\mathbf{H}^c$ & $c^{\text{th}}$ block column from $\mathbf{H}$\\
		$\mathbf{h}[j]$  & $j^{\text{th}}$ entry from $\mathbf{h}$\\
		$\mathcal{S}_i$ & $i^{\text{th}}$ element from set $\mathcal{S}$\\
		$\mathcal{S}^j$ & $j^{\text{th}}$ set\\
		$\mathbf{I}_{L\times L}$ & Identity matrix of size $L\times L$\\
		$\mathbf{r}^{(t)}$ & $\mathbf{r}$ at $t^{\text{th}}$ iteration\\
		$\mathbf{0}_L$ & a length-$L$ vector with all entries equal to $0$\\
		$n^c_{j,i}$ & $i^{\text{th}}$ event from source $c$ in $j^{\text{th}}$ window\\
		$\ast$ , $\star$  & convolution / cross-correlation\\
		$\lVert \cdot\rVert_p$ & $\ell_p$ norm\\
		\\[-0.6em]
		\hline
	\end{tabular}	
	\label{table:notations}
\end{table}

\subsection{Continuous and discrete-time generative models}
Let $y(t)$ be an observed continuous signal in interval $(0,T]$ and $\{h_c(t)\}_{c=1}^C$ be templates (filters) from $C$ sources. We assume that the templates each have the same length and are localized in time. The shift-invariant continuous generative model expresses $y(t)$ as follows
\vspace{-1ex}
\begin{equation}\label{eq:continuous}
y(t)=\sum_{c=1}^C \sum_{i=1}^{N_c} x_{i}^ch_c(t-\tau_{i}^c)+\varepsilon(t),
\end{equation}
where $N_c$ denotes the number of events from source $c$, $\tau_{i}^c$ and $x_{i}^c$ denote the position and the amplitude of the $i^{\text{th}}$ event from source $c$, respectively. The variable $\varepsilon(t)$ denotes i.i.d.~white noise. 

To formulate a discrete-time analogue of Eq.~\ref{eq:continuous}, let $\Delta$ denote the length of a sampling interval and  $f_s=\frac{1}{\Delta}$  the associated sampling frequency. The number of intervals of size $\Delta$ in $(0,T]$ is $N = \lfloor \frac{T}{\Delta}\rfloor$. Further let $n = 1,\cdots,N \in \mathbb{N}^{+}$ be the discrete-time index, and $n_i^c$ denote the discrete-time approximation of $\tau_i^c$, which is such that $n_i^c\Delta\leq \tau_i^c < (n_i^c+1)\Delta$. Finally, we denote by $\mathbf{h}_c\in\mathbb{R}^L$ the discrete time analogue of $h_c(t)$, which we assume is normalized such that $\lVert \mathbf{h}_c\rVert_2=1,\forall c$. Using this notation, we can obtain discrete-time samples $y[n]=y(n\Delta)$ of $y(t)$ that satisfy
\vspace{-1ex}
\begin{equation}
\mathbf{y}[n] = \sum_{c=1}^C \big(\mathbf{x}_{i}^c \ast \mathbf{h}_c\big)[n]+\pmb{\varepsilon}[n],
\label{eq:dtmodel}
\end{equation}
\noindent where we refer to $\mathbf{x}^{c}[n] = \sum_{i=1}^{N_c} x_{i}^c\delta[n-n_{i}^c]$, for $n=1,\cdots,N-L+1$, as \emph{the code}, and $\mathbf{x}^c = \big[\mathbf{x}^c[1],\cdots, \mathbf{x}^c[N-L+1]\big]^{\text{T}}$ as the \emph{code vector}.

\begin{figure}
	\centering
	\includegraphics[width=\linewidth]{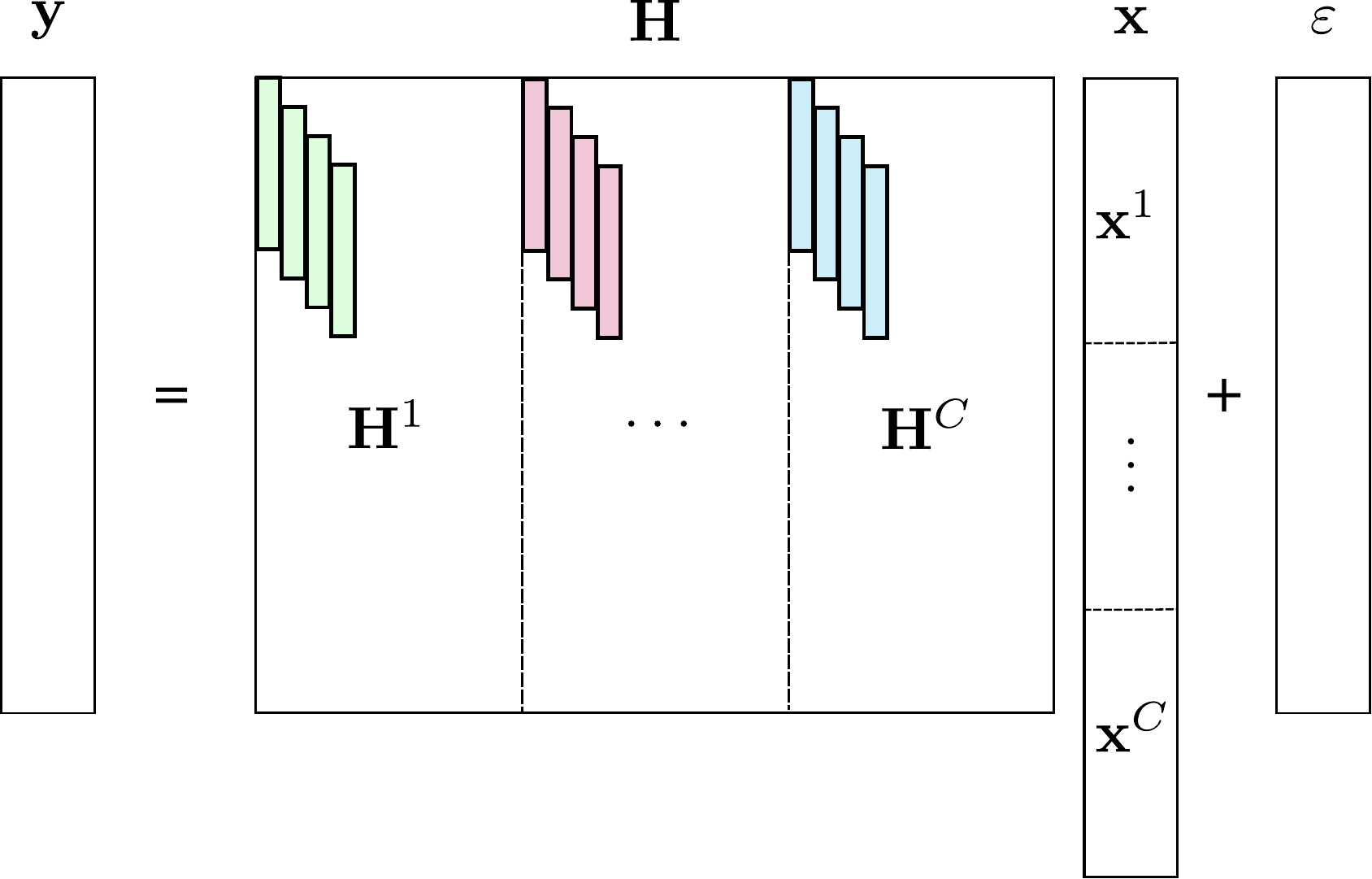}
	\caption{A schematic of Eq. \ref{eq:dtmodel}. The Toeplitz matrix $\mathbf{H}^c$ represents all possible time-shifts of $\mathbf{h}_c$ with integer delay. The non-zero elements of each block $\mathbf{x}^c$ from $\mathbf{x}$ are the times when source $c$ appears in the signal $\mathbf{y}$, $c=1,\cdots,C$.} 
	\label{fig:convolutional_dict}
\end{figure}

We can express Eq.~\ref{eq:dtmodel} in linear-algebraic form as follows
\begin{equation}\label{eq:l0objective}
\mathbf{y}=\mathbf{H}\mathbf{x}+\pmb{\varepsilon},
\end{equation}  
where $\mathbf{H}=\begin{bmatrix}
\mathbf{H}^1\Big\vert \cdots\Big\vert \mathbf{H}^C
\end{bmatrix}\in \mathbb{R}^{N\times C(N-L+1)}$ is a block-Toeplitz matrix with $c^{\text{th}}$ block $\mathbf{H}^c\in \mathbb{R}^{N\times(N-L+1)}$ for $c=1,\cdots,C$ and $\mathbf{x}=[(\mathbf{x}^1)^{\text{T}},\cdots,(\mathbf{x}^C)^{\text{T}}]^{\text{T}}\in\mathbb{R}^{C(N-L+1)}$. The columns of the Toeplitz matrix $\mathbf{H}^c$ represent delayed versions (time-shifts) of $\mathbf{h}_c$, with integer delay between $0$ and $N-L$, that have been zero-padded to have length $N$. For each $c$, the non-zero entries of $\mathbf{x}^c$ represent the discrete-time indices $\{n_i^c\}_{i=1}^{N_c}$ when source $c$ appears in the signal $y[n]$. Fig.~\ref{fig:convolutional_dict} illustrates Eq.~\ref{eq:l0objective}.

In practice, we divide the signal $\mathbf{y}$ into $J$ non-overlapping windows, each of length $W$ and such that $N=WJ$. We assume that $L<<W<<N$, so that the filters from each source are localized within the signal $y[n]$ and within a window. We denote by $\mathbf{Y}\in \mathbb{R}^{W \times J}$ the matrix whose $j^{\text{th}}$ column is $\mathbf{Y}_j = \big[\mathbf{y}[(j-1)W+1],\cdots, \mathbf{y}[jW]\big]^{\text{T}} \in \mathbb{R}^W$, namely the $j^{\text{th}}$ window from $y[n]$. Similarly, we denote by $\mathbf{X}\in\mathbb{R}^{C(W-L+1)\times J}$ the coefficient matrix whose $j^{\text{th}}$ column $\mathbf{X}_{j}=[(\mathbf{x}_{j}^1)^T,\cdots,(\mathbf{x}_{j}^C)^{\text{T}}]^{\text{T}}\in\mathbb{R}^{C(W-L+1)}$ is the code vector associated with window $j$.

\subsection{Optimization objective}
The goal of \textit{Convolutional Dictionary Learning} (CDL) is to estimate $\{\mathbf{h}_c\}_{c=1}^C$ and $\{\mathbf{X}_j\}_{j=1}^J$ that minimize the error of reconstructing the $\mathbf{Y}_j$ in each window using its linear approximation $\mathbf{H}\mathbf{X}_j$. We impose a sparsity constraint on the total number of nonzero elements of $\{\mathbf{X}_j\}_{j=1}^J$ for two reasons. First, without additional constraints, the problem as posed leads to an under-determined system of equations that does not have a unique solution. Second, in many applications, the rate of occurrence of events from the the sources of interest is small compared to $T$, implying that each block $\mathbf{x}_j^c$ of the vector $\mathbf{X}_j$ is sparse. For example, in electrophysiological recordings of neural activity, we expect a sparse number the action potentials from neurons due to their biophysical properties~\cite{Lewicki}. Following~\cite{ksvd}, we use the $\ell_0$ quasi-norm $\lVert \mathbf{X}_j\rVert_0$, which counts the number of non-zero elements of a vector, to express the sparsity constraint. This lets us express the problem mathematically as follows
\vspace{-1ex}
\begin{equation}\label{eq:objective}
\begin{split}
&\min_{\{\mathbf{h}_c\}_{c=1}^C,\{\mathbf{X}_{j}\}_{j=1}^{J}} \sum_{j=1}^J\Big\lVert \mathbf{Y}_j-\mathbf{H}\mathbf{X}_j\Big\rVert_2^2\text{ s.t. } \lVert \mathbf{X}_j\rVert_0\leq \beta_1,\\
\end{split}
\end{equation}  
where $\beta_1$ is a pre-defined sparsity threshold. We refer to this optimization as the CDL problem. One limitation of this approach comes from approximating the continuous-domain generative model (Eq. \ref{eq:continuous}) with the discrete-domain generative model (Eq. \ref{eq:dtmodel}). This approximation results in time-quantization errors, which manifest themselves in two ways: 1) a mismatch between the time when the event occurs in continuous time and its approximation in discrete time,  $\tau^c_{j,i}\neq n^c_{j,i}\Delta$ and 2) template mismatch, $\mathbf{h}_c[m-n^c_{j,i}]=h_c(m\Delta-n^c_{j,i}\Delta)\neq h_c(m\Delta-\tau^c_{j,i})$. One of our contributions is to introduce, in Section \ref{section:comp}, a discrete-time generative model that mitigates the effects of time-quantization errors. 

CDL is a nonconvex optimization problem, due to the simultaneous optimization over $\{\mathbf{h}_c\}_{c=1}^C$ and $\{\mathbf{X}_j\}_{j=1}^J$, as well as the $\ell_0$ penalty. A popular approach is to alternatively minimize the objective over one set of variables while the other is fixed, until convergence. At iteration $t+1$, $\mathbf{X}^{(t+1)}$ is computed based on $\mathbf{H}^{(t)}$ through a \textit{sparse coding} step, after which $\mathbf{H}^{(t+1)}$ is computed using $\mathbf{X}^{(t+1)}$ through a \textit{dictionary update} step. If $\mathbf{H}$ is a convolutional matrix, we refer to these steps as \textit{Convolutional Sparse Coding} (CSC) and \textit{Convolutional Dictionary Update} (CDU), respectively. Existing CSC approaches fall into two categories, which differ with respect to how they enforce the sparsity constraint. One class of approaches, which is the one we follow in this work, uses greedy methods to solve the original problem with the $\ell_0$ quasi-norm. Another class of approaches relaxes the $\ell_0$ quasi-norm to the $\ell_1$ norm, which converts the CSC objective into a convex optimization problem \cite{cardona2017convolutional}, \cite{Chen2001}, \cite{beck2009fast}. The advantage of greedy approaches is that they are more efficient computationally~\cite{TROP2007}. Existing CSC frameworks that address time-quantization errors use the $\ell_1$ norm to enforce sparsity~\cite{CBP}. In the next section, we review both classes of CSC approaches, as well as approaches to solve the CDU step.
 
\section{Background for CDL}\label{section:background}   

For notational simplicity, we use $\mathbf{x}$ instead of $\mathbf{X}_j$ and $\mathbf{y}$  instead of $\mathbf{Y}_j$.

\subsection{Convolutional Greedy Pursuit}

Matching Pursuit (MP) \cite{Mallat1993} and Orthogonal Matching Pursuit (OMP) \cite{Davis1997} are greedy methods to solve the CSC step. We introduce the methods first when $\mathbf{H}$ is an arbitrary matrix, and then discuss the convolutional case.

\subsubsection{Classical greedy pursuit - MP and OMP} Both MP and OMP iteratively select columns from $\mathbf{H}$ to produce an approximation $\mathbf{H}\mathbf{x}$ of $\mathbf{y}$. At iteration step $t'+1$, the column of $\mathbf{H}$ with the maximal absolute inner product with the residual $\mathbf{r}^{(t')}$ is selected and added to the active filter index set $\mathcal{S}^{(t')}$. The initial conditions are $\mathbf{r}^{(0)} = \mathbf{y}$ and $\mathcal{S}^{(0)} = \{\emptyset\}$. The two methods differ in how the coefficients of the chosen columns and the residuals are computed. Let $\mathbf{h}^{(t')}$ denote the template chosen at iteration $t'$.

\begin{itemize}
	\item \textbf{MP} The coefficient associated with $\mathbf{h}^{(t')}$ and the residual are given, respectively, by $\langle \mathbf{h}^{(t')}, \mathbf{r}^{(t')}\rangle$ and $\mathbf{r}^{(t'+1)} = \mathbf{r}^{(t')} - \langle \mathbf{h}^{(t')}, \mathbf{r}^{(t')}\rangle \mathbf{h}^{(t')}$.  
	 
	\item \textbf{OMP} The coefficients associated with $\mathbf{h}^{(1)},\cdots \mathbf{h}^{(t')}$ are those that minimize the squared error between $\mathbf{y}$ and its linear reconstruction using the columns, i.e. those obtained by least-squares. This is equivalent to projecting $\mathbf{y}$ onto the span of $\mathbf{h}^{(1)},\cdots,\mathbf{h}^{(t')}$, and is called a projection step.    
\end{itemize}
The projection step implies two key differences between MP and OMP. First, OMP is slower than MP, due to the matrix inversion and multiplication required in the former. Second, as the residual $\mathbf{r}^{(t')}$ in OMP is orthogonal to the span of previously selected columns, a different column of $\mathbf{H}$ is selected at every iteration. This is not the case for MP, which means that the same column can be selected multiple times.


\subsubsection{Convolutional extensions}
CMP has enjoyed popularity in biomedical applications \cite{spyking, kilosort} and image recognition \cite{Lecun2010}. Convolutional OMP, however, has been less popular, primarily because the projection step with a convolutional matrix is computationally expensive. A previous work \cite{locOMP} suggests a way to improve efficiency, at the expense of accuracy, by projecting the residual onto a subset of the active columns rather than all of them. 

\subsection{CSC for estimating continuous time-shifts}
\begin{figure}[!ht]
	\centering
	\includegraphics[width=\linewidth]{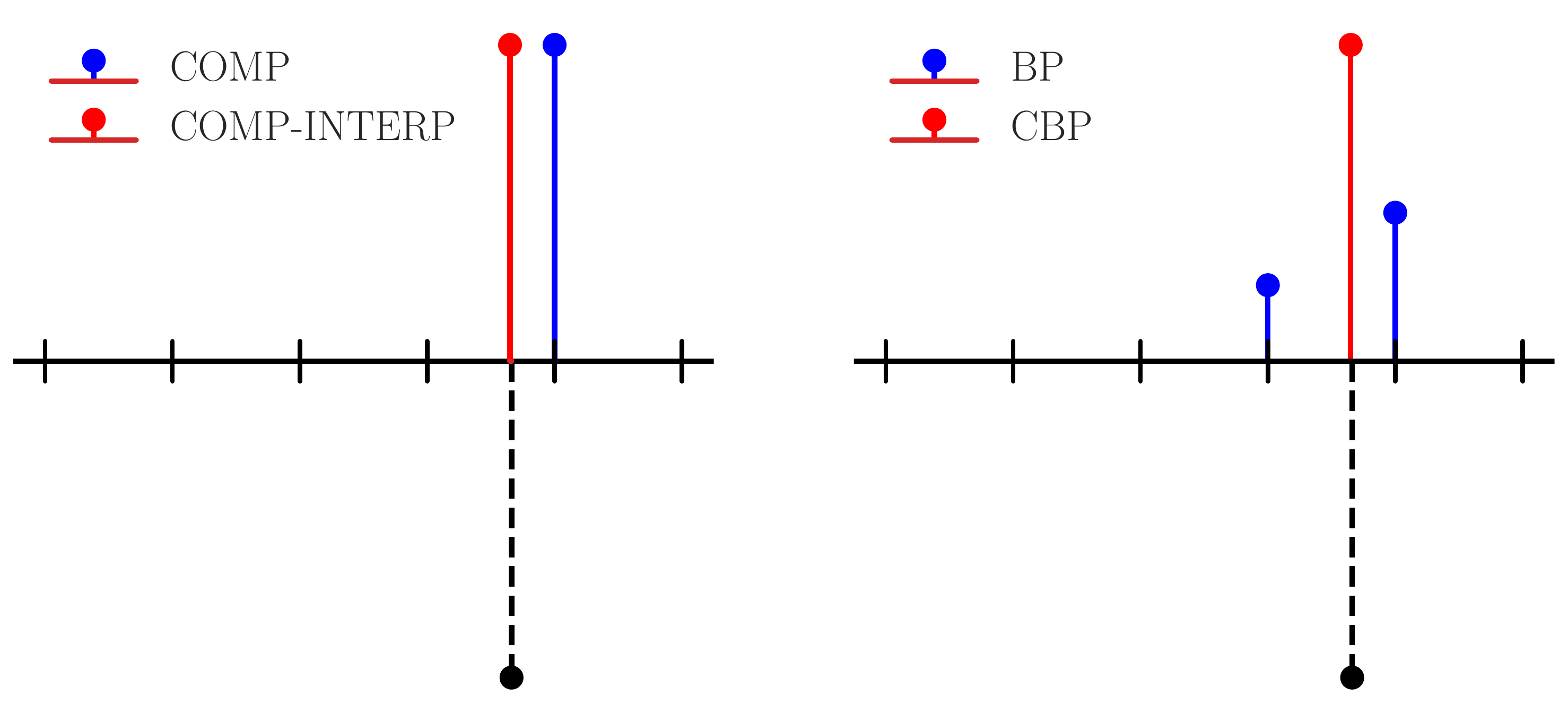}
	\caption{An application of several CSC methods, where an event (black, flipped for clarity) occurs off the discrete-time sampling grid. COMP and BP can only approximate the time of occurrence of the event on the grid. Both CBP and COMP-INTERP (COMP with interpolated dictionary) recover the time of occurrence of the event accurately. COMP-INTERP is much faster than CBP.}
	\label{fig:csc_concept}
\end{figure}

\noindent\textbf{Continuous Basis Pursuit (CBP)} CBP is a CSC framework that uses the $\ell_1$ norm to enforce sparsity, and addresses time-quantization errors~\cite{CBP}. Let $\tau_i^c\in[(m-\frac{1}{2})\Delta,(m+\frac{1}{2})\Delta)$ with $m\in\mathbb{N}$, and $h_c\big(t-\tau_i^c\big)$ be a continuous-time shift of $h_c(t)$. CBP first uses a local basis, in combination with integer shifts $\{h_c(t-m'\Delta)\}_{m'}$ of $h_c(t)$ around $\tau_i^c$, to define a new set of functions $\{\pmb{\phi}^c_p\}_{p=1}^P\in\mathbb{R}^L$ whose linear combination approximates the continuous-time shift. Both $P$ and the number of integer shifts $m'$ depend on the choice of basis. 
A popular option is the Polar basis, where the set of approximating functions are derived from trigonometric splines, with $P=3$ and $m'\in [m,m\pm 1]$. Let $\{\mathbf{\Phi}^{c,p}\}_{p=1}^P\in \mathbb{R}^{N\times (N-L+1)}$ be the convolutional dictionary constructed from the basis and $\mathbf{x}^c_p\in\mathbb{R}^{N-L+1}$ a sparse code vector associated with $\mathbf{\Phi}^{c,p}$. Then, it solves the convex $\ell_1$-regularized optimization problem
\begin{equation}\label{eq:CBP}
\begin{split}
&\min_{\{\vec{\mathbf{x}}^c_i\}} \Big\lVert \mathbf{y} - \sum_{c=1}^C\sum_{p=1}^P\mathbf{\Phi}^{c,p} \mathbf{x}^c_p \Big\rVert_2^2+\lambda\sum_{c=1}^C\lVert\mathbf{x}^c_1 \rVert_1\\
&s.t.\,\,  \big[\mathbf{x}^c_{1}[i],\cdots,\mathbf{x}^c_{P}[i]\big]\in \mathcal{H} \text{ for } i=1,\cdots,N-L+1,\\
\end{split}
\end{equation}
where $\lVert\mathbf{x}^c_1 \rVert_1$ is the $\ell_1$ norm of the amplitude of the first basis element and  $\mathcal{H}$ is convex constraint set that depends on $\{\pmb{\phi}_p^c \}_{c,p=1}^{C,P}$. Lastly, the code $\{\mathbf{x}^c_p \}_{c,p=1}^{C,P}$ is converted to coefficients for the original basis $\{\mathbf{h}_c\}_{c=1}^C$ through an inverse mapping, which yield the times when the events occur and their amplitudes. Fig. \ref{fig:csc_concept} shows an application of CBP, BP, COMP and COMP-INTERP (COMP with interpolated dictionary) to the estimation of the continuous-time shift and amplitude of a single event from one filter ($C=1$). CBP and COMP-INTERP, the approach we propose in Section~\ref{section:comp}, are able to estimate the continuous-time shift accurately. As we demonstrate in Section~\ref{section:experiments}, COMP-INTERP is orders of magnitude faster than CBP. BP and COMP cannot capture the continuous-time shift.

\noindent\textbf{Continuous OMP} In contrast with CBP that uses the $\ell_1$ norm, Continuous OMP~\cite{Knudson2014} is a greedy method to solve Eq.~\ref{eq:CBP}. It uses an extension of OMP to deal with the $\ell_0$ norm. Being greedy, continuous OMP is faster than CBP. Its downside is that it requires the solution to a large number of optimization problems both in the selection and the projection step. 

\subsection{CDU frameworks}
The majority of existing CDU frameworks estimate the templates $\{\mathbf{h}_c\}_{c=1}^C$ by minimizing the error of reconstructing $\mathbf{y}$ using its linear approximation $\mathbf{H}\mathbf{x}$. The key differences between existing approaches are the constraints imposed on the templates and the optimization methods used, as detailed in a recent survey~\cite{cardona2017convolutional}. To the best of our knowledge, existing CDU approaches do not address the problem of learning the templates in the presence of time-quantization errors.

\section{Convolutional Orthogonal Matching Pursuit with Interpolated Dictionary}\label{section:comp}
For the CSC step in the alternating-minimization approach to CDL, we introduce an algorithm for off-the-grid sparse coding called Convolutional OMP (COMP) with interpolated dictionary (COMP-INTERP). This is a convolutional greedy pursuit method that minimizes time-quantization errors. From a computational perspective, COMP-INTERP is an efficient alternative to the $\ell_1$-based CSC frameworks, such as CBP. 

We use $\mathbf{x}$ and $\mathbf{y}$, instead of $\mathbf{X}_j$ and $\mathbf{Y}_j$, for notational simplicity. We use $t$ and $t'$ to denote alternating-minimization iteration and COMP inner iteration, respectively. Since the discussion in this section involves a single iteration of alternating-minimization procedure, we drop $t$.

\begin{figure*}[!ht]
	\centering
	\includegraphics[width=\linewidth]{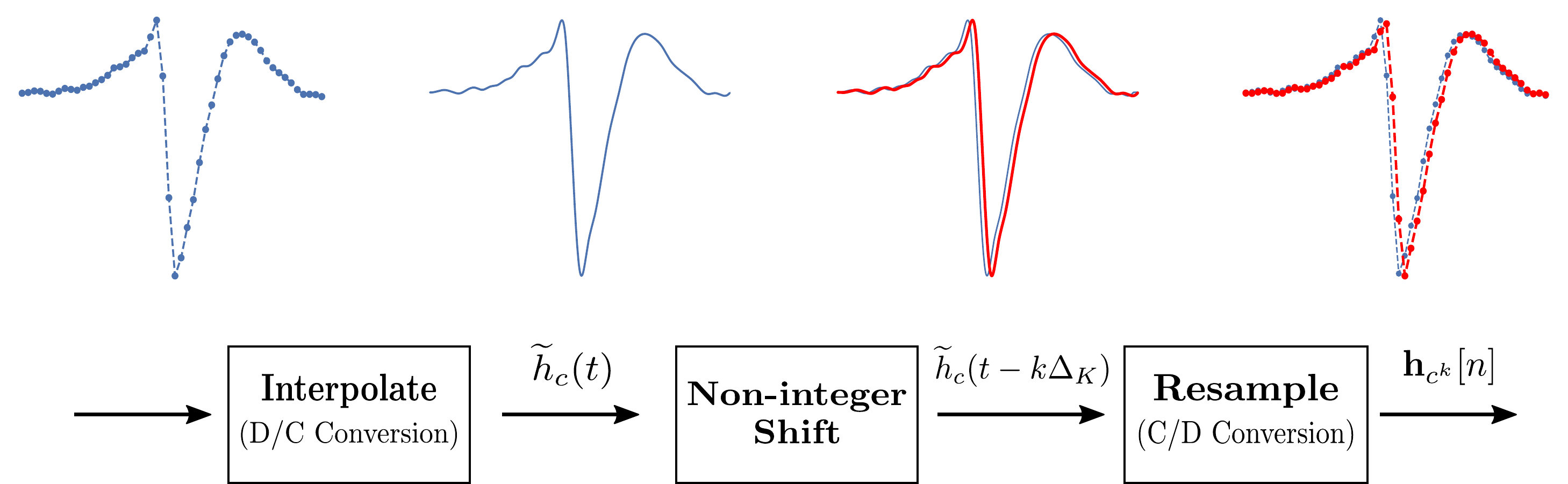}
	\caption{Illustration of the process for obtaining $\mathbf{h}_{c^k}$ (red) from $\mathbf{h}_{c}$ (blue). The two discrete-time templates in the rightmost panel highlight the difference between $\mathbf{h}_{c^k}$ and $\mathbf{h}_{c}$. The Interpolation and Resampling steps correspond, respectively, to D/C (Discrete-to-Continuous conversion and C/D (Continuous-to-Discrete) conversion in digital signal processing theory.}
	\label{fig:interpolation}
\end{figure*}

\subsection{Non-integer delay through smooth interpolation}

The discrete-time model from Eq. \ref{eq:dtmodel} is restrictive because events from the continuous-time generative model of Eq. \ref{eq:continuous} do not necessarily occur at multiples of the  sampling interval $\Delta$. We address this limitation by partitioning $\Delta$ into finer intervals of length $\Delta_K:=\frac{1}{K}\Delta$, and modifying $\mathbf{H}$ and $\mathbf{x}$ accordingly. The resulting CSC framework, with finer resolution $\Delta_K$, can approximate the time of occurrence of the event $\tau_i^c$ with $m\Delta+k\Delta_K$, where $m\in\mathbb{N}$ and $k=0,\cdots,K-1$, rather than with $m\Delta$, which leads to reduction of time quantization error. That is, $\vert \tau_i^c-m\Delta\vert \geq \vert \tau_i^c-(m\Delta+k\Delta_K)\vert$. By definition, each template $\mathbf{h}_c$ corresponds to discrete-time samples of $h_c(t)$ with resolution $\Delta$. Our challenge is to modify $\mathbf{H}$ to account for versions of $h_c(t)$ delayed by a non-integer amount $k\Delta_K$, that is not an integer multiple of $\Delta$, \emph{and} sampled at resolution $\Delta$.
Let $\mathbf{h}_{c^k}\in\mathbb{R}^L$ denote $h_c(t)$ delayed by a non-integer amount $k\Delta_K$ \emph{and} sampled at resolution $\Delta$. This definition motivates us to reformulate Eq.~\ref{eq:dtmodel} to account for non-integer shifts of the templates at a finer scale $\Delta_K$
\vspace{-1ex}
\begin{equation}\label{eq:discrete_modified}
\mathbf{y}[n] = \sum_{c=1}^C \sum_{k=0}^{K-1} \big(\mathbf{x^{c^k}} \ast \mathbf{h}_{c^k}\big)[n]+\pmb{\varepsilon}[n],
\end{equation}

\noindent where $\mathbf{x}^{c^k}$ denotes the code vector corresponding to $\mathbf{h}_{c^k}$. For notational simplicity, we let $\mathbf{h}_{c^0}=\mathbf{h}_c$. Note that $\mathbf{h}_c\neq \mathbf{h}_{c^k}$ for $k\neq 0$, as illustrated in Fig. \ref{fig:interpolation}. The systematic method of obtaining $\mathbf{h}_{c^k}$ from $\mathbf{h}_c$ will be discussed in the next section. We use $\{\mathbf{h}_{c^k}\}_{c,k=1}^{C,K}$ to construct the interpolated convolutional dictionary $\mathbf{\widetilde{H}}\in \mathbb{R}^{W\times CK(W-L+1)}$
\begin{equation}
\mathbf{\widetilde{H}}=\begin{bmatrix}
\mathbf{\widetilde{H}}^{1^0}\Big\vert\cdots \Big\vert \mathbf{\widetilde{H}}^{1^{K-1}}\Big\vert\cdots \Big\vert \mathbf{\widetilde{H}}^{C^0}\Big\vert\cdots \Big\vert \mathbf{\widetilde{H}}^{C^{K-1}}
\end{bmatrix},
\end{equation}
where $\mathbf{\widetilde{H}}^{c^k}$ is the Toeplitz matrix whose columns consists of all integer shifts of $\mathbf{h}_{c^k}$. Note that when $K=1$, we get the original convolutional dictionary, i.e. $\widetilde{\mathbf{H}}=\mathbf{H}$. In linear-algebraic form, we can write the generative model as
$\mathbf{Y}=\mathbf{\widetilde{H}}\mathbf{X}+\pmb{\varepsilon}$, where $\mathbf{X}\in\mathbb{R}^{CK(W-L+1)\times J}$.

\subsection{Smooth interpolation of $\mathbf{h}_c$}\label{section:interpolation}

We use the concept of \textit{continuous-time operations on discrete-time signals} from digital signal processing theory~\cite{Oppenheim2009} to obtain $\mathbf{h}_{c^k}$ from $\mathbf{h}_{c}$. The process consists of three steps: interpolation, shifting by a non-integer amount, and resampling. These steps, illustrated in Fig.~\ref{fig:interpolation}, perform the following operations
\begin{enumerate}
	\item \textbf{Interpolation} Interpolate $\mathbf{h}_c$ with a smooth interpolator to obtain $\tilde{h}_c(t)$.
	\item \textbf{Non-integer shift} Shift $\tilde{h}_c(t)$ to obtain $\tilde{h}_c(t-k\Delta_K)$.
	\item \textbf{Resampling} Resample $\tilde{h}_c(t-k\Delta_K)$ with resolution $\Delta$ to obtain $\mathbf{h}_{c^k}$.
\end{enumerate}

The interpolated template $\tilde{h}_c(t)$ depends on the choice of interpolator and does not necessarily coincide with the continuous-time template $h_c(t)$. In the next section, we show that a sinc interpolator yields $\tilde{h}_c(t)=h_c(t)$.

\subsection{Bandlimited interpolation of $\mathbf{h}_c$}

Suppose $h_c(t)$ is \emph{bandlimited} and that the sampling frequency $f_s$ is above its Nyquist rate. The sinc interpolator then guarantees that $\tilde{h}_c(t)=h_c(t)$, namely that we can recover the continuous-time filter from its discrete-time samples. Approximating a continuous-time signal with discrete-time samples in this manner is commonly referred to as \textit{bandlimited interpolation} \cite{Oppenheim2009}. The intuition behind bandlimited interpolation is two-fold: 1) convolving a signal with a sinc interpolator is equivalent to multiplying the signal with an ideal low-pass filter with unit gain in the frequency domain, and thus the signal is not distorted upon interpolation, and 2) when bandlimited $h_c(t)$ is sampled at $f_s$, there is no aliasing of high frequency content into low frequency bands. The bandlimited assumption translates to the smoothness on $h_c(t)$, which requires that the template does not change abruptly between consecutive samples on the sampling grid. Interpolating $\mathbf{h}_c$ with a sinc interpolator $\text{sinc}(t) = \frac{\sin \pi t}{\pi t}$ yields 
\vspace{-1ex}
\begin{equation}\label{eq:dc}
\begin{split}
\tilde{h}_c(t) &= \mathbf{h}_c*\Big(\frac{\sin \pi t}{\pi t}\Big)\\ 
&=\sum_{m=-\infty}^{m=\infty}\mathbf{h}_c[m]\frac{\sin \pi(t-m\Delta)}{\pi(t-m\Delta)}.
\end{split}
\end{equation}
\noindent To obtain $\mathbf{h}_{c^k}$, we shift $\tilde{h}_c(t)=h_c(t)$ by $k\Delta_K$ and resample with resolution $\Delta$, where $\mathbf{f}^k$ is defined as the discrete sinc interpolator shifted by $k\Delta_K$,
\begin{equation}\label{eq:sinc_shift}
\begin{split}
\mathbf{h}_{c^k}[n]
&=h_c(t-k\Delta_K)\Big\vert_{t=n\Delta}\\
&=\sum_{m=-\infty}^{m=\infty}\mathbf{h}_c[m]\frac{\sin \pi((n-m)\Delta-k\Delta_K)}{\pi((n-m)\Delta-k\Delta_K)}\\
&=\mathbf{h}_c*\underbrace{\Bigg(\frac{\sin \pi(n\Delta-k\Delta_K)}{\pi(n\Delta-k\Delta_K)}\Bigg)}_{\mathbf{f}^k}.\\
\end{split}
\end{equation}

The three-step procedure described previously, and illustrated in Fig.~\ref{fig:interpolation}, 
is equivalent to convolving $\mathbf{h}_c$ with $\mathbf{f}^k$. In practice, since the sinc interpolator has infinite support, we use a truncated sinc interpolator to ensure that $\mathbf{h}_{c^k}$ is of finite length. The truncation provides a good approximation to the infinite discrete sinc interpolator \cite{Oppenheim2009}. Let $\mathbf{f}^k\in \mathbb{R}^L$ be the sinc interpolator restricted to the support $\big[-\frac{L-1}{2},\frac{L-1}{2}\big]$. We obtain the truncated $\mathbf{h}_{c^k}$ by expressing the convolution as a matrix multiplication: $\mathbf{h}_{c^k}=\mathbf{F}^k\mathbf{h}_c$, where $\mathbf{F}^k\in\mathbb{R}^{L\times L}$ is defined as 
\begin{equation}
\mathbf{F}^k=\begin{pmatrix}
\mathbf{f}^k[0] & \mathbf{f}^k[-1] & \cdots & \mathbf{f}^k[-\frac{L-1}{2}] & \mathbf{0}_{\big(\frac{L-1}{2}\big)}^T\\
\mathbf{f}^k[1] & \mathbf{f}^k[0] & \cdots & \mathbf{f}^k[-\frac{L-1}{2}] & \mathbf{0}_{\big(\frac{L-3}{2}\big)}^T\\
& &\vdots & &\\
\mathbf{f}^k[\frac{L-1}{2}] &  & \cdots &  & \mathbf{f}^k[-\frac{L-1}{2}]\\
& &\vdots & &\\
\mathbf{0}_{\big(\frac{L-1}{2}\big)}^T & \mathbf{f}^k[\frac{L-1}{2}] & \cdots & & \mathbf{f}^k[0] \\
\end{pmatrix}.
\end{equation}
The matrix $\mathbf{F}^k$ is the Toeplitz (convolutional) matrix associated with $\mathbf{f}^k$, truncated so that it is of $L\times L$. The truncation ensures that the resulting $\mathbf{h}_{c^k}$ is of the same length as $\mathbf{h}_c$.

\subsection{Efficient algorithm for COMP and COMP-INTERP}
Matrix operations involving convolutional dictionaries $\mathbf{H}$ or $\widetilde{\mathbf{H}}$ are expensive both in terms of computations and storage requirements. This is because typical recordings can last on the order of minutes, if not hours, and sampling rates can be on the order of $\sim10^4 (Hz)$ for electrophysiology and $10^3 (Hz)$ for EEG, to name a few examples. Existing greedy algorithms for CSC can handle high-dimensional data in the selection step because it involves cross-correlation operations 
for which efficient and scalable implementations exist. They fall short, however, of addressing the inefficiency of the projection step in high-dimensional settings. This has led to the wide adoption of CMP and not COMP.

We explore efficient implementations of the selection and the projection step for a convolutional matrix $\mathbf{H}$ (or $\widetilde{\mathbf{H}}$), which lead to efficient implementations of COMP and COMP-INTERP.
An efficient implementation of the selection step, which we restate for completeness, was proposed in~\cite{locOMP}. For notational simplicity, we focus on $\mathbf{H}$, noting that the same argument holds for $\mathbf{\widetilde{H}}$. \\

\noindent\textbf{Selection step}
The selection step requires the inner product between time-shifted $\mathbf{h}_c$ and $\mathbf{r}^{(t')}$, expressed as $\mathbf{H}^{\text{T}}\mathbf{r}^{(t')}$. For large $\mathbf{H}$, explicit computation of the inner product is expensive. However, the convolutional structure of $\mathbf{H}$ lets compute $C$ cross-correlations instead \cite{Zhang2008MLSSL}
\begin{equation}\label{eq:cross_correlation}
\begin{split}
\mathbf{H}^{\text{T}} \mathbf{r}^{(t')}=&\big[(\mathbf{h}_1\star \mathbf{r}^{(t')})[1], \cdots,(\mathbf{h}_1\star \mathbf{r}^{(t')})[N-L+1],\cdots,\\
&(\mathbf{h}_C\star \mathbf{r}^{(t')})[1],\cdots,(\mathbf{h}_C\star \mathbf{r}^{(t')})[N-L+1]\big]^{\text{T}},
\end{split}
\end{equation}
where the cross-correlation index $i$ indicates the time offset of $\mathbf{h}_c[1]$ relative to the $\mathbf{r}^{(t')}[1]$. This formulation has two benefits:
\begin{enumerate}
	\item We do not need to construct the convolutional matrix $\mathbf{H}$ explicitly. We only require $O(N)$ memory to store $\{\mathbf{h}_c\}_{c=1}^C$ and $\mathbf{r}^{(t')}$, as opposed to $O(CN^2)$ memory to store the matrix. 
	\item We can compute the $C$ cross-correlation operations using the FFT, which is much more efficient than computing them by multiplication of $\mathbf{H}^{\text{T}}$ and $\mathbf{r}^{(t')}$.
\end{enumerate}

\noindent\textbf{Projection step}
In this step, we project the residual onto the span of $\mathbf{H}\big|_{t'}$, which requires the inversion of $\mathbf{H}\big|_{t'}^{\text{T}}\mathbf{H}\big|_{t'}$. The matrix $\mathbf{H}\big|_{t'}\in \mathbb{R}^{N\times t'}$ refers to a convolutional dictionary restricted to columns that have been selected by COMP up to $t'^{\text{th}}$ iteration. Consequently, the code $\mathbf{x}|_{t'}\in \mathbb{R}^{t'}$ refers to the nonzero coefficients from $\mathbf{x}\in \mathbb{R}^{N-L+1}$ corresponding to the columns $\mathbf{H}\big|_{t'}$. 
To avoid the computational cost of inversion in the projection step,  ~\cite{rubinstein2008efficient} suggested an efficient method for computing the Cholesky factor $\mathbf{L}^{(t')}$, which is a lower triangular matrix such that $\mathbf{L}^{(t')}\big(\mathbf{L}^{(t')}\big)^{\text{T}}=\mathbf{H}\big|_{t'}^{\text{T}}\mathbf{H}\big|_{t'}\in\mathbb{R}^{t'\times t'}$. The key idea is that for OMP, $\mathbf{H}\big|_{t'-1}$ and $\mathbf{H}\big|_{t'}$ differ only by one column, which is the column selected by OMP at step $t'$, and therefore $\mathbf{L}^{(t')}$ can be easily computed from $\mathbf{L}^{(t'-1)}$ as
\vspace{-1ex}
\begin{equation}\label{eq:cholesky}
\begin{split}
&\mathbf{L}^{(t')}=\begin{pmatrix}
\mathbf{L}^{(t'-1)} & 0 \\
\mathbf{w}^{\text{T}} & \sqrt{1-\lVert \mathbf{w}\rVert_2^2}\\
\end{pmatrix}\\
&\text {where } \mathbf{L}^{(t'-1)}\mathbf{w}=\mathbf{H}\big|_{t'-1}^{\text{T}}\mathbf{h}^{(t')},\\
\end{split}
\end{equation}
where $\mathbf{h}^{(t')}\in\mathbb{R}^{N}$ denotes the column of $\mathbf{H}$ selected at iteration $t'$ of COMP. The code $\mathbf{x}\big|_{t'}$ is obtained as the solution to  $\mathbf{L}^{(t')}\big(\mathbf{L}^{(t')}\big)^{\text{T}}\mathbf{x}|_{t'}=\mathbf{H}\big|_{t'}^{\text{T}}\mathbf{y}$, which can be solved more efficiently than the linear system $\mathbf{H}\big|_{t'}^{\text{T}}\mathbf{H}\big|_{t'}\mathbf{x}|_{t'}=\mathbf{H}\big|_{t'}^{\text{T}}\mathbf{y}$. 

We extend this idea to the convolutional case, noting that Eq.~\ref{eq:cholesky} still requires us to construct $\mathbf{H}\big|_{t'}$ and to perform multiplications that are expensive in terms of memory and computation. We replace the multiplication operation involving $\mathbf{H}\big|_{t'}^{\text{T}}$ with a cross-correlation operation, as outlined in Algorithm \ref{alg:cholesky}. To keep track of the selected filters, we utilize two sets: 1) The set $\mathcal{S}$ of active template indices chosen by COMP and defined in Section \ref{section:background}, and 2) the set $\mathcal{I}$ of times when events associated with each of the templates from $\mathcal{S}$ occur. At COMP iteration $t'$, $\mathbf{h}_{\mathcal{S}_i^{(t')}}$ refers to the template selected at $i^{\text{th}}$ iteration, where $i\leq t$, and $\mathcal{I}^{(t')}_i$ refers to the time of occurrence of the corresponding template. 
\begin{algorithm}[!ht]
	\DontPrintSemicolon
	\KwIn{$\mathbf{L}^{(t'-1)}$, $\mathcal{S}^{(t')}$, $\mathcal{I}^{(t')}$, $\{\mathbf{h}_c\}_{c=1}^C$, $\mathbf{y}$}
	\KwOut{ $\mathbf{L}^{(t')}$, $\mathbf{x}|_{t'}$}
	Initialization: $\mathbf{v}\in \mathbb{R}^{t'-1}$, $\pmb{\alpha} \in \mathbb{R}^{N}$\;
	\For{$i\leftarrow 1$ \KwTo $t'-1$}{
		\eIf{$\Big|\mathcal{I}^{(t')}_i-\mathcal{I}^{(t')}_{t'}\Big|>L$}{
			$\mathbf{v}[i]=0$}{
			$\mathbf{v}[i]=\Big(\mathbf{h}_{\mathcal{S}^{(t')}_i} \star \mathbf{h}_{\mathcal{S}^{(t')}_{t'}}\Big)\Big[\mathcal{I}^{(t')}_i-\mathcal{I}^{(t')}_{t'}\Big]$
		}		
	}
	solve for $\mathbf{w}$: $\mathbf{L}^{(t'-1)}\mathbf{w}=\mathbf{v}$\;
	$\mathbf{L}^{(t')}\leftarrow \begin{pmatrix}
	\mathbf{L}^{(t'-1)} & 0\\
	\mathbf{w}^{\text{T}} & \sqrt{1-\lVert \mathbf{w}\rVert_2^2}\\
	\end{pmatrix}$\;
	\For{$i\leftarrow 1$ \KwTo $t'$}{
		$\pmb{\alpha}[i]=(\mathbf{h}_{\mathcal{S}^{(t')}_i}\star \mathbf{y})\Big[\mathcal{I}^{(t')}_i\Big]$\;
	}
	Solve for $\mathbf{L}^{(t')}\big(\mathbf{L}^{(t')}\big)^{\text{T}}\mathbf{x}|_{t'}=\pmb{\alpha}$\;
	\caption{Cholesky factorization for CSC at $t'$}\label{alg:cholesky}
\end{algorithm}
With $\mathcal{S}$ and $\mathcal{I}$, neither the convolutional matrix $\mathbf{H}\big|_{t'}$, nor the zero-padded filters are required, which makes COMP very efficient. We emphasize that the efficiency gain from the modified projection step outweighs that of the selection step, as we demonstrate in Section \ref{section:experiments}. Specifically, the efficient projection step essentially removes the huge performance gap between the CMP and the COMP.\\

\noindent\textbf{Algorithm}
We summarize the implementation of the efficient COMP-INTERP in Algorithm \ref{alg:comp}. When $K=1$, COMP-INTERP is equivalent to COMP with the original, non-interpolated, dictionary. The INTERPOLATE function refers to the process of obtaining interpolated templates. The CHOLESKY function refers to the efficient projection step. The superscript $(\ast)$ denotes the quantities at convergence. As in OMP, the convergence criterion can either be when the residual falls below a certain threshold or when a certain sparsity level is reached.
\begin{algorithm}[!ht]
	\DontPrintSemicolon
	\SetKwFunction{Interpolate}{Interpolate}
	\SetKwFunction{CHOLESKY}{CHOLESKY}
	\SetKwFunction{ZeroProjection}{ZeroProjection}
	\SetKwFunction{NNCD}{NNCD}
	\KwIn{$\mathbf{y}$, $\{\mathbf{h}_c\}_{c=1}^C$, $\frac{\Delta}{K}$}
	\KwOut{$\mathbf{x}^{(\ast)}, \mathcal{S}^{(\ast)}, \mathcal{I}^{(\ast)}$}
	Initialization: $\mathbf{r}^{(0)}=\mathbf{y}$, $\mathbf{L}^{(0)}=1$, $\mathcal{S}^{(0)}, \mathcal{I}^{(0)}=\{\emptyset\}$\;
	$\{\mathbf{h}_{c^k} \}_{c,k=1}^{C,K}\leftarrow$\Interpolate{$\{\mathbf{h}_c\}_{c=1}^C$, $\frac{\Delta}{K}$}\;
	\While{$t'=0$ \KwTo convergence}{
		(\textit{Selection step})\;
		$(c^k)^{\ast},i^{\ast}\leftarrow \arg\max_{c,k,i}\{\mathbf{h}_{c^k}\star\mathbf{r}^{(t')}[i]\}_{c,k,i=1}^{C,K,N-L+1}$\;
		$\mathcal{S}^{(t'+1)}=\mathcal{S}^{(t')}\cup \{(c^k)^{\ast}\}$,   $\mathcal{I}^{(t'+1)}=\mathcal{I}^{(t')}\cup \{i^{\ast}\}$\;
		(\textit{Projection step})\;
		$\mathbf{L}^{(t'+1)},\mathbf{x}|_{t'+1}\leftarrow$\CHOLESKY{$\theta_{\text{CHOL}}$}, where\;
		$\quad\theta_{\text{CHOL}}=\big\{\mathbf{L}^{(t')}, \mathcal{S}^{(t'+1)}, \mathcal{I}^{(t'+1)}, \{\mathbf{h}_{c^k}\}_{c,k=1}^{C,K}, \mathbf{y}\big\}$\;
		$\mathbf{r}^{(t'+1)}\leftarrow \mathbf{y}$\;
		
		\For{$i\leftarrow 1$ \KwTo $t'+1$}{
			Subtract $\mathbf{h}_{\mathcal{S}^{(t'+1)}_i}\mathbf{x}|_{t'+1}$ from the segment of $\mathbf{r}^{(t'+1)}$ that starts at $\mathcal{I}_i^{(t'+1)}$
		}
	}
	\caption{COMP-INTERP}\label{alg:comp}
\end{algorithm}

\subsection{Comparison with the continuous basis approach}
The continuous-basis approximation methods, CBP and continuous OMP, and COMP-INTERP all minimize the effect of time quantization errors due to events off the grid. All three methods assume that the continuous-time templates are smooth. In the continuous-basis approximation approach, the derivation of the mapping that relates the local basis and the templates relies on the smoothness assumption. In COMP-INTERP, the bandlimited assumption implies the smoothness of the templates.   

\begin{figure}
	\centering
	\includegraphics[width=\linewidth]{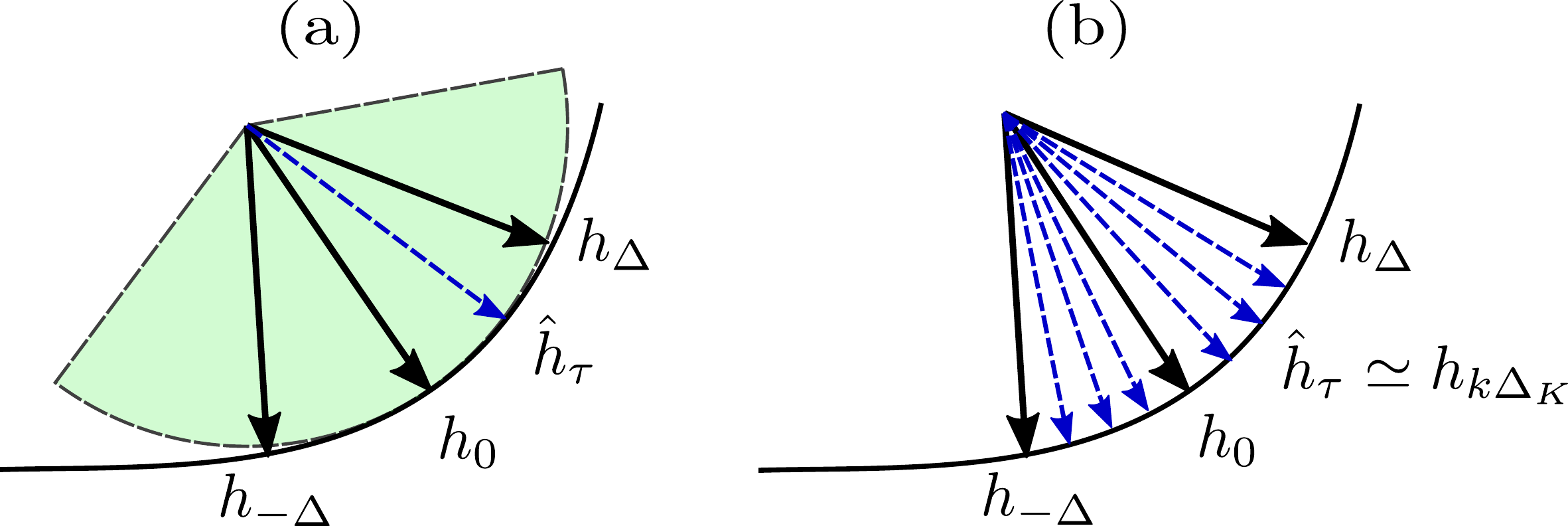}
	\caption{ Illustration of how CBP (polar basis) and COMP-INTERP approximate $\hat{h}_{\tau}\simeq h(t-\tau)$. (a) CBP constructs a circular arc to locally approximate the subspace of continuously-translated $h(t)$, and chooses the mostly likely point on the arc. (b) COMP-INTERP directly places the original template $h(t)$ at the finer resolution $\Delta_K$.}
	\label{fig:CBP_vs_COMP}
\end{figure}

The two approaches differ in how the templates are used, as illustrated in Fig. \ref{fig:CBP_vs_COMP}. As detailed in Section~\ref{section:background}, the continuous-basis approach approximates continuous-time shifts of the templates by using the linear combination of a set of functions indexed by points on the discrete-time sampling grid. COMP-INTERP takes an approach that is simpler and more direct: it uses bandlimited interpolation to approximate continuous-time shifts of the templates by multiples of a arbitrarily fine interval $\Delta_K$. This leads to an overcomplete, interpolated dictionary, comprising the original templates and their interpolated variants. In fact, the dictionary $\mathbf{\widetilde{H}}$, can be thought of as a discrete approximation to the infinite-dimensional dictionary that encodes all possible continuous-time shifts ~\cite{atomic}.

The advantages of our approach are its simplicity and its speed. It is  simpler as it requires neither a local basis nor a mapping between said basis and the templates. It is faster due to its greedy nature, and because of the efficient implementations of operations that involve the convolutional dictionary. 

\section{Convolutional Dictionary Update with interpolated dictionary}\label{section:cdl}
COMP-INTERP is an algorithm to solve the CSC step of CDL using a interpolated dictionary that can approximate continuous-time shifts. In this section, we develop an algorithm to solve the CDU step using the interpolated dictionary. 

The CDU step involves an optimization problem with respect to $\{\mathbf{h}_c\}_{c=1}^C$. To simplify it, we first re-write $\mathbf{H}\mathbf{X}_j$ as
\vspace{-1ex}
\begin{equation}\label{eq:commutativity}
\mathbf{H}\mathbf{X}_j = \sum_{c=1}^C\sum_{i=1}^{N_c}x_{j,i}^c\mathbf{S}_{j,i}^c\mathbf{h}_c,
\end{equation}
where we introduce the matrix representation $\mathbf{S}_{j,i}^c\in\mathbb{R}^{W\times L}$ of the linear operator that shifts  $\mathbf{h}_c$ by $n_{j,i}^c$ samples. $\mathbf{S}_{j,i}^c$ is a zero-padded identity matrix defined as follows 
\begin{equation}
\mathbf{S}_{j,i}^c=
\begin{pmatrix}
	\mathbf{0}_{n_{j,i}^c\times L}\\
	\mathbf{I}_{L\times L}\\
	\mathbf{0}_{(W-L-n_{j,i}^c)\times L}
\end{pmatrix}.  
\end{equation}
Eq.~\ref{eq:commutativity} is a result of the commutativity of convolution operation. 
It allows us to re-write the optimization problem in Eq.~\ref{eq:objective}, with respect to $\{\mathbf{h}_c\}_{c=1}^C$, as follows 
\vspace{-1ex}
\begin{equation}\label{eq:cdl_objective}
\min_{\{\mathbf{h}_c\}_{c=1}^C} \sum_{j=1}^J\Big\lVert \mathbf{Y}_j-\sum_{c=1}^C\sum_{i=1}^{N_c}x_{j,i}^c\mathbf{S}_{j,i}^c\mathbf{h}_c\Big\rVert_2^2.
\end{equation}
Compared to Eq.~\ref{eq:objective}, Eq.~\ref{eq:cdl_objective} is simpler because $\mathbf{h}_c$ appears as a vector, as opposed to a matrix. Using the interpolated dictionary $\widetilde{\mathbf{H}}$ to account for non-integer delays, we can write the objective similarly 
\vspace{-1ex}
\begin{equation}\label{eq:cdl_interpolated_objective}
\begin{split}
&\min_{\{\mathbf{h}_c\}_{c=1}^C} \sum_{j=1}^J\Big\lVert \mathbf{Y}_j-\sum_{c=1}^C\sum_{k=1}^K\sum_{i=1}^{N_{c^k}}x_{j,i}^{c^k}\mathbf{S}_{j,i}^{c^k}\mathbf{h}_{c^k}\Big\rVert_2^2\\
&\text{s.t. }\mathbf{h}_{c^k}=\mathbf{F}^k\mathbf{h}_c,\quad\forall k, c,\\
\end{split}
\end{equation}
where $\mathbf{F}^k$ is the sinc interpolator matrix defined previously. The constraint enforces the fact we obtain $\mathbf{h}_{c^k}$ from $\mathbf{h}_{c}$ by interpolation, as discussed in Section \ref{section:comp}. As is customary in the CDL literature, we assume the templates have unit norm, a constraint that we enforce by normalizing the solutions of Eq. \ref{eq:cdl_objective} or Eq. \ref{eq:cdl_interpolated_objective}.

\subsection{CDU with basic dictionary}
To solve Eq. \ref{eq:cdl_objective}, we can use any least-squares based algorithm for dictionary. We focus on KSVD~\cite{ksvd}, specifically shift-invariant KSVD~\cite{AharonPhD}. Two key ideas from KSVD are worth emphasizing: 1) the templates $\{\mathbf{h}_c\}_{c=1}^C$ are updated sequentially, i.e.~one at a time, and 2) when updating a given template, only windows of the data where the template occurs need to be considered.

Suppose we want to update template $\mathbf{h}_{\overline{c}}$. Let $\Omega_{\overline{c}}\subset \{1,\cdots,J\}$ denote the set of indices of windows from which COMP selects at least one occurrence of $\mathbf{h}_{\overline{c}}$. For each window $j\in \Omega_{\overline{c}}$, we split the sum from Eq.~\ref{eq:cdl_objective} into two parts, namely one that involves $\mathbf{h}_{\overline{c}}$ and another that involves the remaining templates. The new estimate of $\mathbf{h}_{\overline{c}}$, denoted by $\widehat{\mathbf{h}}_{\overline{c}}$ is given by
\vspace{-1ex}
\begin{equation}\label{eq:cdl_leastsquares}
\begin{split}
&\widehat{\mathbf{h}}_{\overline{c}}=\arg\min_{\mathbf{h}_{\overline{c}}}\sum_{j\in \Omega_{\overline{c}}}\Bigg\lVert \mathbf{E}_j -\sum_{i=1}^{N_{\overline{c}}}x_{j,i}^{\overline{c}}\mathbf{S}_{j,i}^{\overline{c}}\mathbf{h}_{\overline{c}}\Bigg\rVert_2^2,\\
&\text{where }\mathbf{E}_j=\mathbf{Y}_j - \sum_{c\neq \overline{c}}^C\sum_{i=1}^{N_c}x_{j,i}^{c}\mathbf{S}_{j,i}^{c}\mathbf{h}_{c},\\
\end{split}
\end{equation}

\noindent is the residual from approximating  $\mathbf{Y}_j$ with templates other than $\mathbf{h}_{\overline{c}}$. Eq.~ \ref{eq:cdl_leastsquares} is a least-squares problem whose solution $\widehat{\mathbf{h}}_{\overline{c}}$ is given by
\vspace{-1ex}
\begin{equation}\label{eq:cdl_solution}
\begin{split}
\widehat{\mathbf{h}}_{\overline{c}} &= \Bigg(\sum_{j\in\Omega_{\overline{c}}}\sum_{i=1}^{N_{\overline{c}}}\sum_{m=1}^{N_{\overline{c}}}x_{j,i}^{\overline{c}}\Big(\mathbf{S}_{j,i}^{\overline{c}}\Big)^{\text{T}}\mathbf{S}_{j,m}^{\overline{c}}x_{j,m}^{\overline{c}}\Bigg)^{-1}\\
&\times\Bigg(\sum_{j\in\Omega_{\overline{c}}}\sum_{i=1}^{N_{\overline{c}}}x_{j,i}^{\overline{c}}\Big(\mathbf{S}_{j,i}^{\overline{c}}\Big)^{\text{T}}\mathbf{E}_j\Bigg).\\
\end{split}
\end{equation}
We can interpret Eq.~\ref{eq:cdl_solution} as the weighted average of segments $\mathbf{E}_j$, or $\big(\mathbf{S}_{j,i}^{\overline{c}}\big)^{\text{T}}\mathbf{E}_j\in\mathbb{R}^L$ in which template $\mathbf{h}_{\overline{c}}$ occurs. The average is normalized by a factor that accounts for occurrences of $\mathbf{h}_{\overline{c}}$ that overlap: the term $\big(\mathbf{S}_{j,i}^{\overline{c}}\big)^{\text{T}}\mathbf{S}_{j,m}^{\overline{c}}\in \mathbb{R}^{L\times L}$ is a matrix that is non-zero only if the offset $\vert n_{j,i}^{\overline{c}}-n_{j,m}^{\overline{c}}\vert$ between occurrences of $\mathbf{h}_{\overline{c}}$ is less than the template length $L$. 

\subsection{CDU with interpolated dictionary (CDU-INTERP)}

To solve Eq.~\ref{eq:cdl_interpolated_objective}, we follow an approach similar to that used to solve Eq.~\ref{eq:cdl_objective}. The constraint from Eq.~\ref{eq:cdl_interpolated_objective} implies that 
\vspace{-1ex}
\begin{equation}\label{eq:interpolation_manipulation}
\begin{split}
\mathbf{S}_{j,i}^{c^k}\mathbf{h}_{c^k}=\mathbf{S}_{j,i}^{c^k}\mathbf{F}^k\mathbf{h}_{c}=\widetilde{\mathbf{S}}_{j,i}^{c^k}\mathbf{h}_{c},
\end{split}
\end{equation}
where $\widetilde{\mathbf{S}}_{j,i}^{c^k}:=\mathbf{S}_{j,i}^{c^k}\mathbf{F}^k$ is a modified shift operator (with $\mathbf{F}^k$, instead of $\mathbf{I}_{L\times L}$, placed in the block starting from row $n_{j,i}^{c^k}$). Eq.~\ref{eq:interpolation_manipulation} allows us to rewrite Eq.~\ref{eq:cdl_interpolated_objective} as the unconstrained optimization problem
\vspace{-1.5ex}
\begin{equation}\label{eq:cdl_interpolated_objective2}
\min_{\{\mathbf{h}_c\}_{c=1}^C} \sum_{j=1}^J\Big\lVert \mathbf{Y}_j-\sum_{c=1}^C\sum_{k=1}^K\sum_{i=1}^{N_{c^k}}x_{j,i}^{c^k}\widetilde{\mathbf{S}}_{j,i}^{c^k}\mathbf{h}_{c}\Big\rVert_2^2.
\end{equation}

\noindent Suppose we want to update template $\mathbf{h}_{\overline{c}}$. Let $\widetilde{\Omega}_{\overline{c}}=\cup_{k=1}^K\Omega_{\overline{c}^{k}}$ be the set of indices of windows from which COMP selects at least one occurrence of a template from the set $\{\mathbf{h}_{\overline{c}^{k}}\}_{k=1}^K$. Re-arranging Eq.~\ref{eq:cdl_interpolated_objective2} yields the estimate $\widehat{\mathbf{h}}_{\overline{c}}$ of $\mathbf{h}_{\overline{c}}$
\vspace{-1ex}
\begin{equation}
\begin{split}
&\widehat{\mathbf{h}}_{\overline{c}}=\arg\min_{\mathbf{h}_{\overline{c}}}\sum_{j\in \Omega_{\overline{c}}}\Bigg\lVert \mathbf{E}_j -\sum_{k=1}^K\sum_{i=1}^{N_{\overline{c}^{k}}}x_{j,i}^{\overline{c}^{k}}\widetilde{\mathbf{S}}_{j,i}^{\overline{c}^{k}}\mathbf{h}_{\overline{c}}\Bigg\rVert_2^2\\
&\text{where }\mathbf{E}_j=\mathbf{Y}_j - \sum_{c\neq \overline{c}}^C\sum_{k=1}^K\sum_{i=1}^{N_{c^k}}x_{j,i}^{c^{k}}\widetilde{\mathbf{S}}_{j,i}^{c^{k}}\mathbf{h}_{c},\\
\end{split}
\end{equation}
the solution of which is given by
\vspace{-1ex}
\begin{equation}\label{eq:cdl_interpolated_solution}
\begin{split}
\widehat{\mathbf{h}}_{\overline{c}} &= \Bigg(\sum_{j\in\widetilde{\Omega}_{\overline{c}}}\sum_{k=1}^K\sum_{i=1}^{N_{\overline{c}}}\sum_{m=1}^{N_{\overline{c}}} x_{j,i}^{\overline{c}^k}\Big(\widetilde{\mathbf{S}}_{j,i}^{\overline{c}^k}\Big)^{\text{T}}\widetilde{\mathbf{S}}_{j,m}^{\overline{c}^k}x_{j,m}^{\overline{c}^k}\Bigg)^{-1}\\
&\times\Bigg(\sum_{j\in\widetilde{\Omega}_{\overline{c}}}\sum_{k=1}^K\sum_{i=1}^{N_{\overline{c}}}x_{j,i}^{\overline{c}^k}\Big(\widetilde{\mathbf{S}}_{j,i}^{\overline{c}^k}\Big)^{\text{T}}\mathbf{E}_j\Bigg).\\
\end{split}
\end{equation}

Similar to Eq.~\ref{eq:cdl_solution}, we can interpret Eq.~\ref{eq:cdl_interpolated_solution} as the weighted average of the interpolated segments $\mathbf{E}_j$ in which template any template from the set $\{\mathbf{h}_{\overline{c}^{k}}\}_{k=1}^K$ occurs.



\subsection{CDL algorithm}

We summarize the alternating-minimization procedure for CDL in Algorithm \ref{alg:CDL}. COMP-INTERP refers to the Algorithm \ref{alg:comp} and CDU refers to the CDU step. The $\ast$ notation from $\mathcal{S}^{j,(\ast)}$ and $\mathcal{I}^{j,(\ast)}$ refer to the index sets for window at COMP convergence. 
\begin{algorithm}
	\DontPrintSemicolon
	\SetKwFunction{COMPINTERP}{COMP-INTERP}
	\SetKwFunction{CDU}{CDU}
	\KwIn{$\mathbf{Y},\{\mathbf{h}_c^{(0)}\}_{c=1}^C,\frac{\Delta}{K}$}
	\KwOut{$\mathbf{X}^{(\ast)}$, $\{\mathbf{h}^{(\ast)}_c\}_{c=1}^C$}
	\While{$t=0$ \KwTo convergence}{
		(\textit{CSC step})\;
		\For{$j=1$ \KwTo $J$}{
			$\mathbf{X}_j^{(t+1)},\,
			\mathcal{S}^{j,(\ast)},\,\mathcal{I}^{j,(\ast)}$\; $\leftarrow$ \COMPINTERP\big($\mathbf{Y}_j, \{\mathbf{h}_c^{(t)}\}_{c=1}^C,\frac{\Delta}{K}$\big)}\;
		(\textit{CDU step})\;
		$\{\mathbf{\widehat{h}}_c\}_{c=1}^C\leftarrow \{\mathbf{h}_c^{(t)}\}_{c=1}^C$\;
		\For{$c=1$ \KwTo$ C$}{
			$\mathbf{\widehat{h}}_c$\; 
			$\leftarrow $\CDU{$\mathbf{Y},\{\mathbf{\widehat{h}}_c\}_{c=1}^C$,$\{\mathbf{X}_j^{(t+1)},\,\mathcal{S}^{j,(\ast)},\,\mathcal{I}^{j,(\ast)}\}_{j=1}^J$,$\frac{\Delta}{K}$}
		}
		$\{\mathbf{h}_c^{(t+1)}\}_{c=1}^C\leftarrow \{\mathbf{\widehat{h}}_c\}_{c=1}^C$\;
	}
	\caption{CDL}\label{alg:CDL}
\end{algorithm}
COMP-INTERP is parallelizable across $J$ windows, and therefore amenable to implementation on GPU and parallel computer architectures. The CDU step is not parallelizable  because it needs to aggregate the occurrences of the templates across all $J$ windows.

\section{Experiments}\label{section:experiments}
We apply the proposed CDL framework to simulated data and the real electrophysiological data from the brain. We use two criteria to evaluate performance: 1) the accuracy and speed of the CSC step, and 2) the accuracy of the CDU step. 

For the CSC step, we compare CBP to the following convolutional greedy methods: COMP, CMP, COMP-slow, and COMP-INTERP. CMP does not have a projection step. The projection step from COMP-slow inverts $\mathbf{H}\big|_{t'}^{\text{T}}\mathbf{H}\big|_{t'}$ directly without Cholesky factorization.  Table~\ref{table:methods} summarizes the similarities and differences of the greedy methods.

\begin{table}[!ht]
	\centering
	\caption{Convolutional greedy methods}
	\begin{tabular}{|l|c|c|c|}
		\hline
		& Efficient selection & Efficient projection & $\Delta_K$\\
		\hline
		COMP & Yes & Yes & $\Delta$\\
		CMP & Yes & $\cdot$ & $\Delta$\\
		COMP-slow & Yes & No & $\Delta$\\
		COMP-INTERP & Yes & Yes & $\Delta/K$ \\
		\hline
	\end{tabular}
	\label{table:methods}
\end{table}

\subsection{Simulated Dataset}
\noindent We simulated a signal according to the continuous-time generative model of Eq.~\ref{eq:continuous}. We used two $10$-ms-long gamma-tone templates \cite{CBP} defined for $t$ in the interval from $-5$ to $5$ ms
\begin{equation}
\begin{split}
&h_1(t)\propto \big(10^3t\big)\exp\Big(-\big(10^3t\big)^2\Big)\cos\Big(\frac{\pi}{2}\big(10^3t\big)\Big)\\
&h_2(t)\propto \big(10^3t\big)\exp\Big(-\big(10^3t\big)^2\Big).\\
\end{split} 
\end{equation}
We assumed the same number of occurrences $N_1=N_2$ of the templates. We chose the times when events occur uniformly at random, i.e.~$\tau_i^c\sim \text{Uniform}[0,T]$ for $c=1,2$ and $i=1,\cdots,N_1$. We chose the amplitude of each event uniformly at random, i.e.~$x_i^c\sim \text{Uniform}[1, 2]$. As explained in subsequent sections, we used a range of values for the variables $T$ and $N_1$. 
We used a sampling rate $f_s=10^4$ Hz and obtained the discrete-time signal $\mathbf{y}$ by sampling $y(t)$ at every $\Delta=10^{-4}$ seconds. We added white Gaussian noise $\pmb{\varepsilon} \sim \mathcal{N}(0,\sigma^2\mathbf{I})$ in discrete time, where $\sigma$ was set according to a desired Signal-to-noise ratio (SNR). We defined the SNR as follows
\begin{equation}
\text{SNR}=10\log \left(\frac{\lVert \mathbf{H}\mathbf{x}\rVert_2^2}{\lVert \varepsilon\rVert_2^2}\right)\quad \text{(dB)},
\end{equation} 

\noindent where the variance estimate $\widehat{\sigma}^2$ is $\widehat{\sigma}^2=\frac{\lVert \varepsilon\rVert_2^2}{N}$, with $N = \lfloor \frac{T}{\Delta}\rfloor$. We obtained $\mathbf{h}_1,\mathbf{h}_2\in \mathbb{R}^{100}$ by acquiring 100 samples from $h_1(t)$ and $h_2(t)$ and normalizing the resulting vectors to have unit length: $\lVert\mathbf{h}_1\rVert_2=\lVert\mathbf{h}_2\rVert_2=1$.

\subsection{Results from simulations: CSC step}

\noindent  We set the sparsity level for greedy methods to be the number of events, and fine-tuned the regularization parameter $\lambda$ for CBP (with polar basis) to match the same sparsity level. We use the true templates $\mathbf{h}_1,\mathbf{h}_2$ as dictionary elements.

\noindent\textbf{Sparse-coding computation time} We computed the duration of the CSC step,  using CBP and the above-mentioned greedy methods, as a function of data length $T$ and total number of events $N_1+N_2$. Specifically, we ran two sets of experiments: 1) $T\in[0.5,1,2,3,4,5]\,\text{s}$ with fixed $N_1+N_2=30$ and 2) $N_1+N_2\in[10,20,30,40,50]$ with fixed $T=3$ s. For each experiment, we report durations averaged over 50 independent repeats (trials).
Tables~\ref{table:computation1} and~\ref{table:computation2} show the duration of the CSC step for the various methods, respectively as a function of data length and number of occurrences of events. We draw three conclusions from the results

\begin{enumerate}
	\item  COMP is computationally much faster than CBP, with two possible explanations. In terms of implementation, CBP constructs the full convolutional dictionary for $\ell_1$ regularization, whereas COMP does not. Moreover, it is well-known that greedy methods are faster than ones based on basis-pursuit like methods~\cite{TROP2007},~\cite{Kunis2008}.  
	
	\item COMP is as fast as CMP. This is true even for large $T$ or $N_1+N_2$, which involve a computationally-demanding projection step. This, along with a comparison of COMP to COMP-slow highlights the importance of making the projection step efficient. 
	 COMP reduces the computation time of COMP-slow by 48\% $\sim$ 85\% on average.
		
	\item The difference between COMP and COMP-INTERP stems from the fact that the selection step of the former requires $C$ cross-correlation operations, while that of the latter requires $KC$ such operations. The computation time of the projection step is the same for both.
\end{enumerate}

\begin{table}[!ht]
	\centering
	\caption{Computation time (sec) as a function of $T$}
	\begin{tabular}{|l|c|c|c|c|c|c|}
		\hline
		& 0.5 & 1 & 2 & 3 & 4 & 5\\
		\hline
		\textbf{COMP} & \textbf{0.041} & \textbf{0.066} & \textbf{0.114} & \textbf{0.175} & \textbf{0.223} & \textbf{0.271} \\
		CMP & 0.027 & 0.049 & 0.089 & 0.131 & 0.180 & 0.237 \\
		COMP-slow & 0.079 & 0.176 & 0.350 & 0.545 & 0.759 & 1.002 \\
		\textbf{COMP-INTERP} & \textbf{0.243} & \textbf{0.445} & \textbf{0.781} & \textbf{1.161} & \textbf{1.565} & \textbf{2.038} \\
		CBP & 4.24 & 13.55 & 51.75 & 145.27 & $\cdot$ & $\cdot$ \\
		\hline
	\end{tabular}
\label{table:computation1}
\end{table}

\begin{table}[!ht]
	\centering
	\caption{Computation time (sec) as a function of $N_1+N_2$}
	\begin{tabular}{|l|c|c|c|c|c|}
		\hline
		& 10 & 20 & 30 & 40 & 50\\
		\hline
		\textbf{COMP} & \textbf{0.096} & \textbf{0.204} & \textbf{0.267} & \textbf{0.340} & \textbf{0.438}\\
		CMP & 0.082 & 0.168 & 0.218 & 0.276 & 0.345\\
		COMP-slow & 0.194 & 0.666 & 1.189 & 2.111 & 3.043\\
		\textbf{COMP-INTERP} & \textbf{0.744} & \textbf{1.416} & \textbf{1.811} & \textbf{2.377} & \textbf{3.112}\\
		CBP & 152.44 & 134.25 & 135.10 & 132.60 & 133.53\\ 	
		\hline
	\end{tabular}
	\label{table:computation2}
\end{table}

\noindent\textbf{Sparse coding accuracy} We computed the \textit{average hit error}~\cite{Knudson2014} for COMP and CBP, and that for COMP-INTERP as a function of the discretization $\Delta_K$. The \textit{average hit error} measures how far, in terms of absolute displacement, the recovered sparse codes are relative to the true sparse codes. We simulated 50 trials of data with $T=1$ and $N_1=N_2=10$ and computed the median average hit error across trials for each method. The average hit error of both COMP and CBP is independent of $\Delta_K$: COMP operates at the sampling resolution $\Delta$, while CBP operates in continuous time.

\begin{figure}[!ht]
	\centering
	\includegraphics[width=\linewidth]{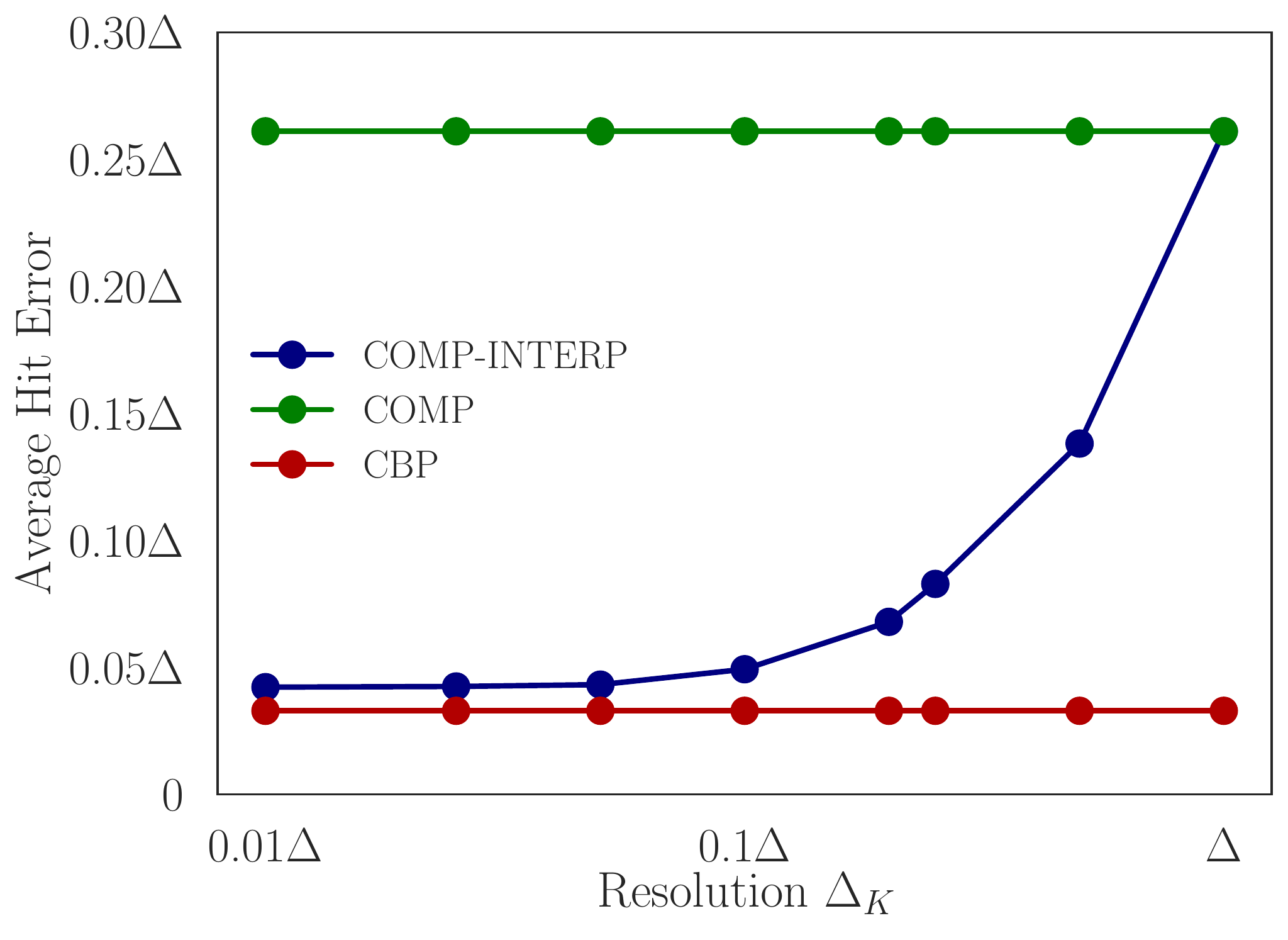}
	\caption{Average hit error as a function of $\Delta_K$, on a log scale, for CBP, COMP, and COMP-INTERP. Each point represents the median \textit{average hit error} computed across 50 trials of simulated data.}
	\label{fig:csc_comparison}
\end{figure}

Fig.~\ref{fig:csc_comparison} shows that the average hit error for COMP-INTERP is lower than that for COMP. The finer the discretization, the greater the precision with which COMP-INTERP identifies the sparse codes, resulting in the lower average hit error. 
The reduction in average hit error is marginal below a certain threshold ($\Delta_{10}=\frac{1}{10}\Delta$). The finer the discretization, the more correlated the interpolated templates and, in turn, the columns of the dictionary are. Therefore, as discretization becomes smaller, the returns, in terms of average hit error, from expanding the dictionary diminish. 
With no discretization, COMP and COMP-INTERP achieve the same average hit errors since the original and the interpolated dictionary coincide. Fig.~\ref{fig:csc_comparison} also shows that there is a small gap, between CBP and COMP-INTERP. We observed in simulations that the difference comes from the fact that COMP-INTERP is not as accurate as CBP in the presence of significant template overlaps. Due to its greedy nature, OMP has worse performance in resolving overlapping templates than approaches based on basis pursuit~\cite{Chen2001},~\cite{BLOOMP2}. This can be mitigated by heuristics \cite{BLOOMP}, which we leave it as a future work. 

\subsection{Results from simulations: CDU step}
\noindent\textbf{Dictionary learning accuracy} To assess the accuracy of the CDU algorithms, we use the following error distance metric ~\cite{altmin} between two templates $\widehat{\mathbf{h}}_c$ and $\widetilde{\mathbf{h}}_c$
	\begin{equation}
	\text{err}(\widehat{\mathbf{h}}_c, \widetilde{\mathbf{h}}_c)=\sqrt{1-\frac{\langle \widehat{\mathbf{h}}_c,\widetilde{\mathbf{h}}_c\rangle^2}{\lVert \widehat{\mathbf{h}}_c\rVert_2^2 \lVert \widetilde{\mathbf{h}}_c\rVert_2^2}}.
	\label{eq:err_met}
	\end{equation}
	The lower the metric, the closer $\widehat{\mathbf{h}}_c$ and $\widetilde{\mathbf{h}}_c$ are. If $\widehat{\mathbf{h}}_c=\mathbf{h}_c^{(t)}$ and $\widetilde{\mathbf{h}}_c=\mathbf{h}_c$, the metric measures how close the learned template at iteration $t$ is to the true template. If $\widehat{\mathbf{h}}_c=\mathbf{h}_c^{(0)}$ and $\widetilde{\mathbf{h}}_c=\mathbf{h}_c^{(\ast)}$, the metric measures how much the template at convergence has changed from the initial template.

We compared the accuracy of the CDU step with the interpolated and the original dictionary. We simulated 25 trials of data, with $T=5$ s and $N_1=N_2=200$, for several levels of SNR ranging from 8 to 26 dB. We obtained the initial templates $\mathbf{h}_1^{(0)}$ and $\mathbf{h}_2^{(0)}$ by perturbing the original templates $\mathbf{h}_1$ and $\mathbf{h}_2$ with Gaussian additive noise, such that $\text{err}(\mathbf{h}_c^{(0)}, \mathbf{h}_c)\geq 0.5$ for $c=1,2$. We performed 15 iterations of the alternating-minimization algorithm for CDL to obtain $\mathbf{h}_1^{(\ast)}$ and $\mathbf{h}_2^{(\ast)}$. We compared the following two methods: 1) COMP \& CDU with discretization $\Delta$ (CDL with $\Delta$), which corresponds to conventional CDU methods that use  $\mathbf{H}$ instead of $\widetilde{\mathbf{H}}$, and 2) COMP-INTERP \& CDU-INTERP with discretization $\Delta_{10}=\frac{1}{10}\Delta$ (CDL with $\Delta_{10}$). 

\begin{figure}
	\centering
	\includegraphics[width=\linewidth]{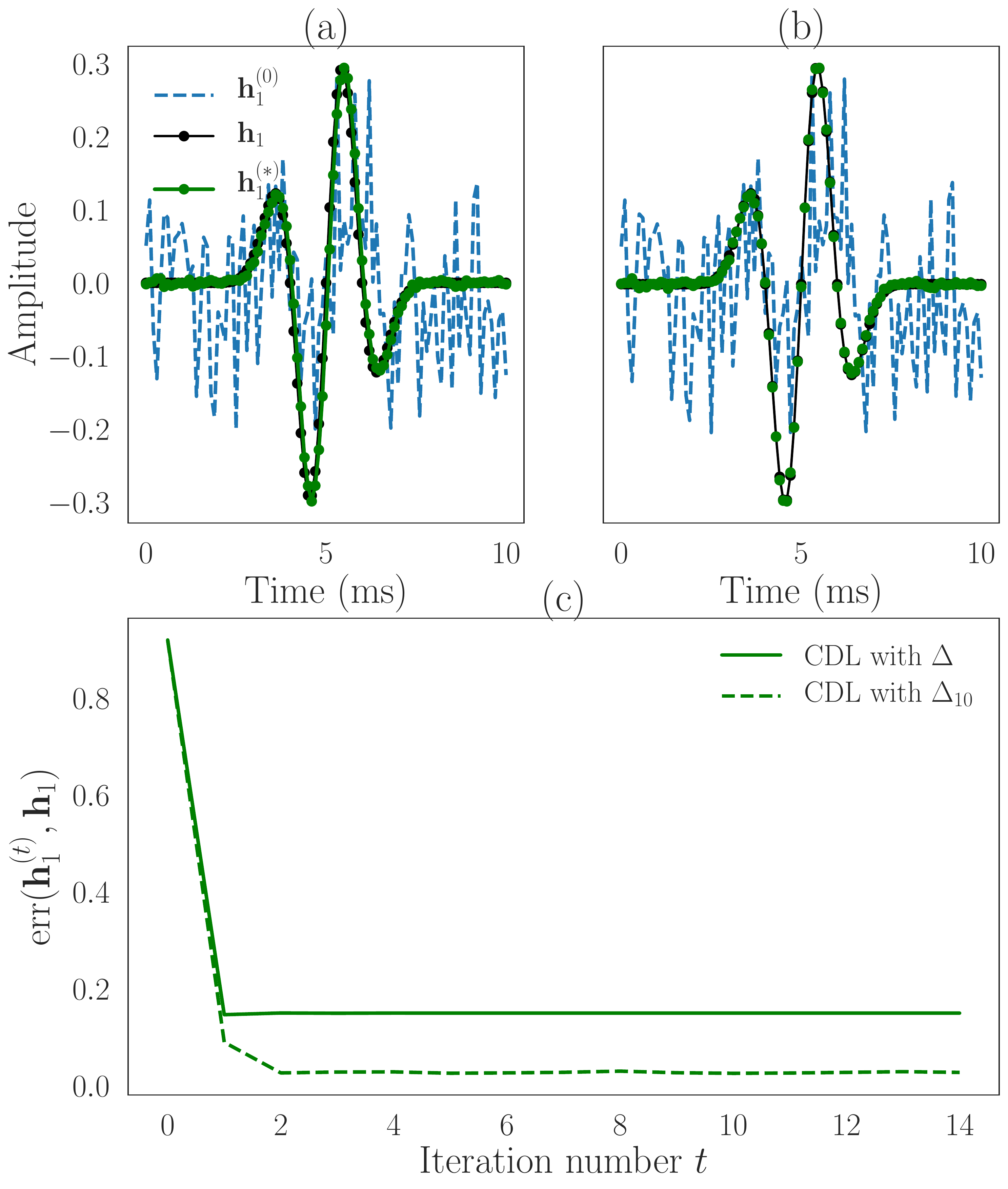}
	\caption{Illustration of the initial $\mathbf{h}_1^{(0)}$, learned $\mathbf{h}_1^{(\ast)}$, and the true template $\mathbf{h}_1$. (a) $\mathbf{h}_1^{(\ast)}$ learned with COMP \& CDU with discretization $\Delta$ (CDL with $\Delta$), (b) learned with COMP-INTERP \& CDU-INTERP with discretization $\Delta_{10}=\frac{1}{10}\Delta$ (CDL with $\Delta_{10}$). The latter approach learns the template more accurately compared to the former. (c) The metric $\text{err}(\mathbf{h}_1^{(t)},\mathbf{h}_1)$ as a function of alternating-minimization iterations.}
	\label{fig:cdl}
\end{figure}

\begin{figure}
	\centering
	\includegraphics[width=\linewidth]{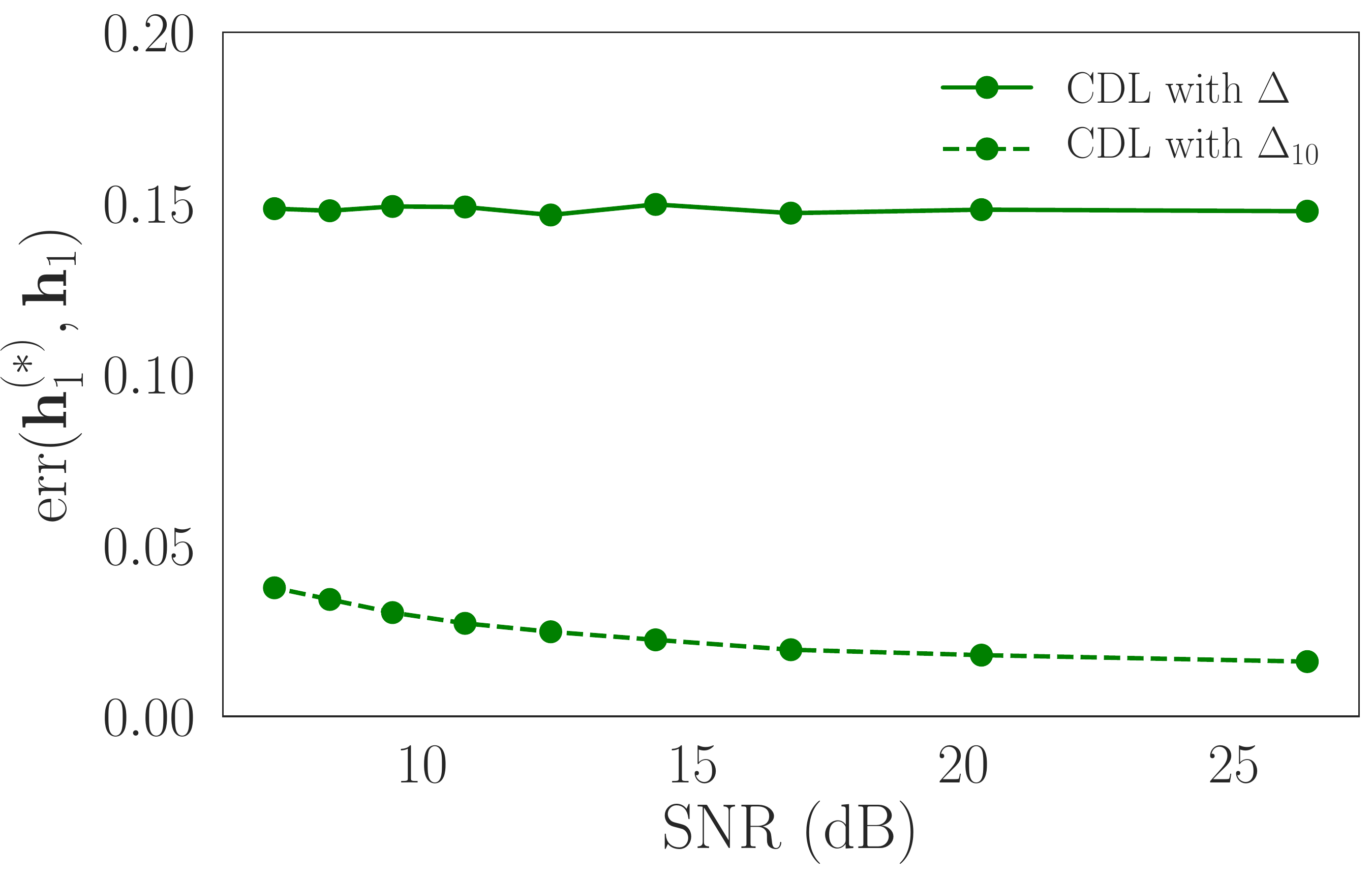}
	\caption{Average, as a function of SNR, of the metric $\text{err}(\mathbf{h}_1^{(\ast)}, \mathbf{h}_1)$ for CDL with $\Delta$ and CDL with $\Delta_{10}$. Each point represents $\text{err}(\mathbf{h}_1^{(\ast)}, \mathbf{h}_1)$ averaged over 25 independent trials.}
	\label{fig:cdl_snr}
\end{figure}

Fig.~\ref{fig:cdl} shows the result of applying the two CDL algorithms to the simulated data. The figure shows that CDL with $\Delta_{10}$ is able to learn templates more accurately than CDL with $\Delta$. This is supported both visual inspection of the learned templates (Fig.~\ref{fig:cdl}(a) and (b)), and the trajectory of $\text{err}(\mathbf{h}_1^{(t)}, \mathbf{h}_1)$ as a function of alternating minimization iteration $t$, for one of the simulated trials. Although not shown here, we observed that CDL with $\Delta_{10}$ learned $\mathbf{h}_2^{(\ast)}$ more accurately than CDL with $\Delta$. CDL with $\Delta$, which does not account account for the non-integer shift, converges to an average of the non-integer shifts of each of the true templates. Fig.~\ref{fig:cdl_snr} shows that both CDL approaches are robust across a range of SNRs.

\subsection{Real dataset: Application to spike sorting}
We applied our framework to spike sorting. Given a recording of extracellular voltage, the goal of spike sorting is to learn the action potentials (templates) from neurons (sources) near the electrode, and the times when the action potentials occur~\cite{Lewicki}. We used a dataset that consists of an extracellular recording from the rat hippocampus, along with a simultaneous intracellular recording \cite{Henze2000} from one neuron. The intracellular recording provides the \textit{ground truth} data, as it provides the unequivocal occurrence of an action potential in a single neuron, and thus enables us to evaluate and compare the accuracy of the CDL frameworks. For this dataset, 620 events occurred from the neuron that was recorded intracellularly. 
The sampling rate of the extracellular data, which comprise 4 channels, is $f_s=10^4$ Hz. We used $T=150$ seconds of data from channel 1 and preprocessed them following the procedures that are standard in the spike-sorting literature~\cite{Simoncelli}. Specifically, we applied to the data a high-pass filter with cut-off frequency 400 Hz, and whitened it. In addition, we identified peaks from the data that crossed a pre-defined threshold \cite{Quiroga} and extracted a segment of length $81$ samples centered around each peak. 
The resulting collection $\mathbf{Y}\in\mathbb{R}^{5000\times 81}$ of $5,000$ segments is the input to our analyses of the real data.   

\noindent \textbf{Method setup} We assumed $C=3$, namely that the extracellular recording can detect activity from 3 neurons. We used templates $\{\mathbf{h}_c\}_{c=1}^3\in \mathbb{R}^{41}$, each of length 4 ms. We applied CDL, with COMP and COMP-INTERP in the CSC step, to the extracellular data. We used the following procedure to initialize the templates to $\big\{ \mathbf{h}_c^{(0)} \big\}_{c=1}^3$. Following segment extraction, we first performed PCA on $\mathbf{Y}$ for dimensionality reduction, and then K-means clustering with three clusters in the lower dimension space. We used the centroids of the clusters to obtain $\big\{ \mathbf{h}_c^{(0)} \big\}_{c=1}^3$. We used an estimate of the variance of the background noise in the recording as the termination criterion for the CSC step of COMP and COMP-INTERP. We computed this estimate by extracting data from a segment that remained below a pre-defined threshold for more than 500 ms. We ran 15 iterations of the CDL algorithm to obtain $\{\mathbf{h}_c^{(\ast)}\}_{c=1}^3$.

We compared CBP with $\{\mathbf{h}_c^{(0)}\}_{c=1}^3$ to the CSC step of COMP and COMP-INTERP using the filters, $\{\mathbf{h}_c^{(\ast)}\}_{c=1}^3$, learned by the respective CDU algorithms. For a given true spike event from the intracellular data, we associate an event identified using the extracellular data as a \textit{true positive} if the event is within 30 samples (3 ms) of a true event from the intracellular data. Among the templates learned using the extracellular data, we associate the template with the highest true positive count with the neuron from the intracellular data and refer to it as $\mathbf{h}_1$. As we do not have access to the true $\mathbf{h}_1$, we treat $\mathbf{h}_1^{(\ast)}$ as the best estimate of the true spike template.

\subsection{Results from the real dataset}
\noindent\textbf{Detection error curve}
We used two statistics to evaluate the ability of the methods to perform spike sorting on the extracellular data. Following CSC, each method must set a threshold to identify the times when action potentials (spikes) occur. A \textit{true miss} is a true spike from the intracellular data within 3 ms of which no threshold-crossing event occurs in the extracellular data. The true miss rate is the ratio of the number of true misses to the number of intracellular events. A \textit{false alarm} is a threshold-crossing event from the extracellular data that is not a true spike, i.e.~within 3 ms of which no events occur in the intracellular data. The false alarm rate is the ratio of the number of false alarms to the number of threshold crossing events. Varying the threshold leads to a trade-off between true misses and false alarms. A threshold with high amplitude typically leads to a low number of false alarms and a large number of true misses, and vice versa for a low-amplitude threshold. 

\begin{figure}
	\centering
	\includegraphics[width=\linewidth]{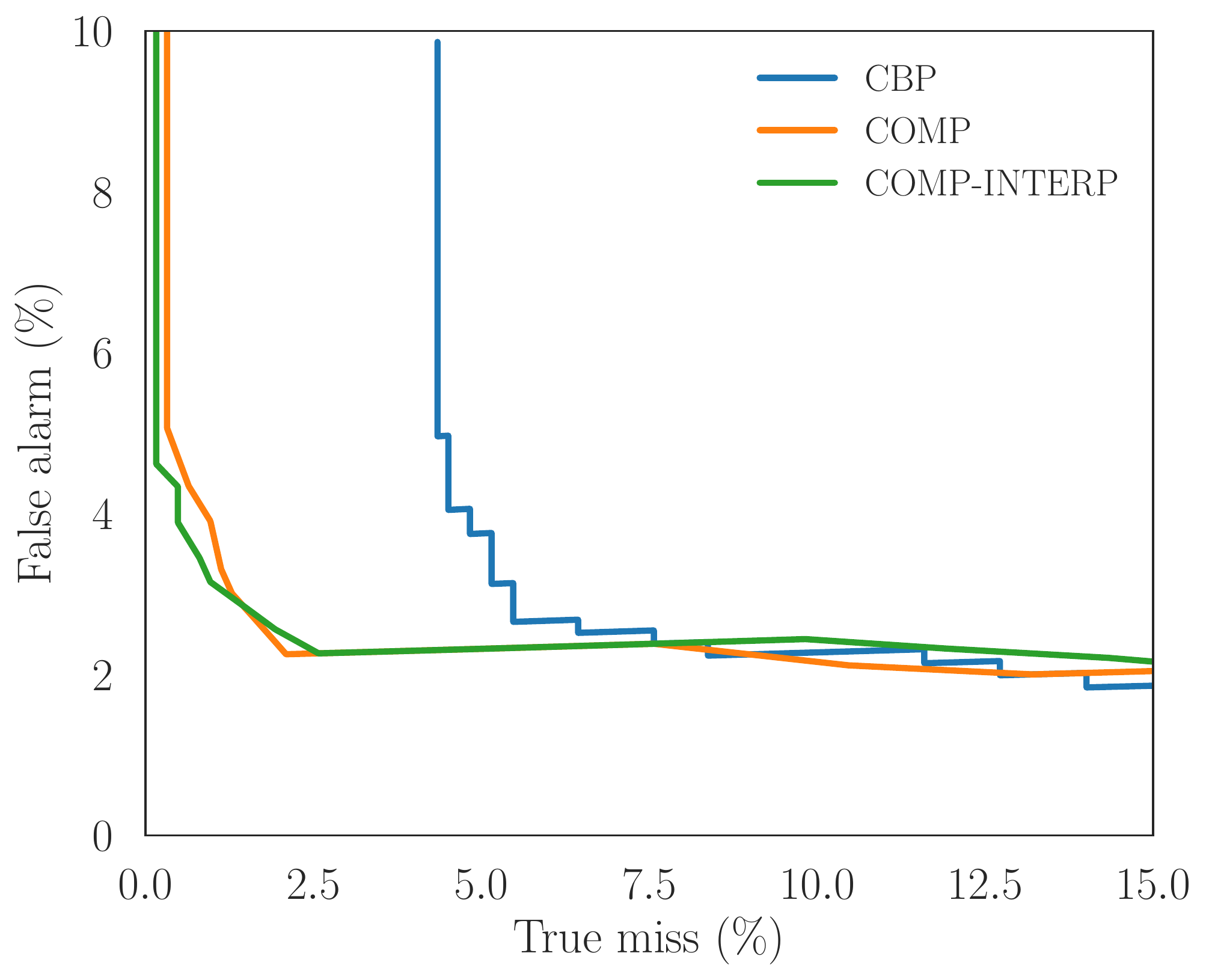}
	\caption{Error curves for events associated with $\mathbf{h}_1$ (true spike template). The curves, computed for CBP, COMP and COMP-INTERP, show the trade-off between the false alarm rate and the true miss rate for each method.}
	\label{fig:real_error_curve}
\end{figure}
\begin{figure}
	\centering
	\includegraphics[width=\linewidth]{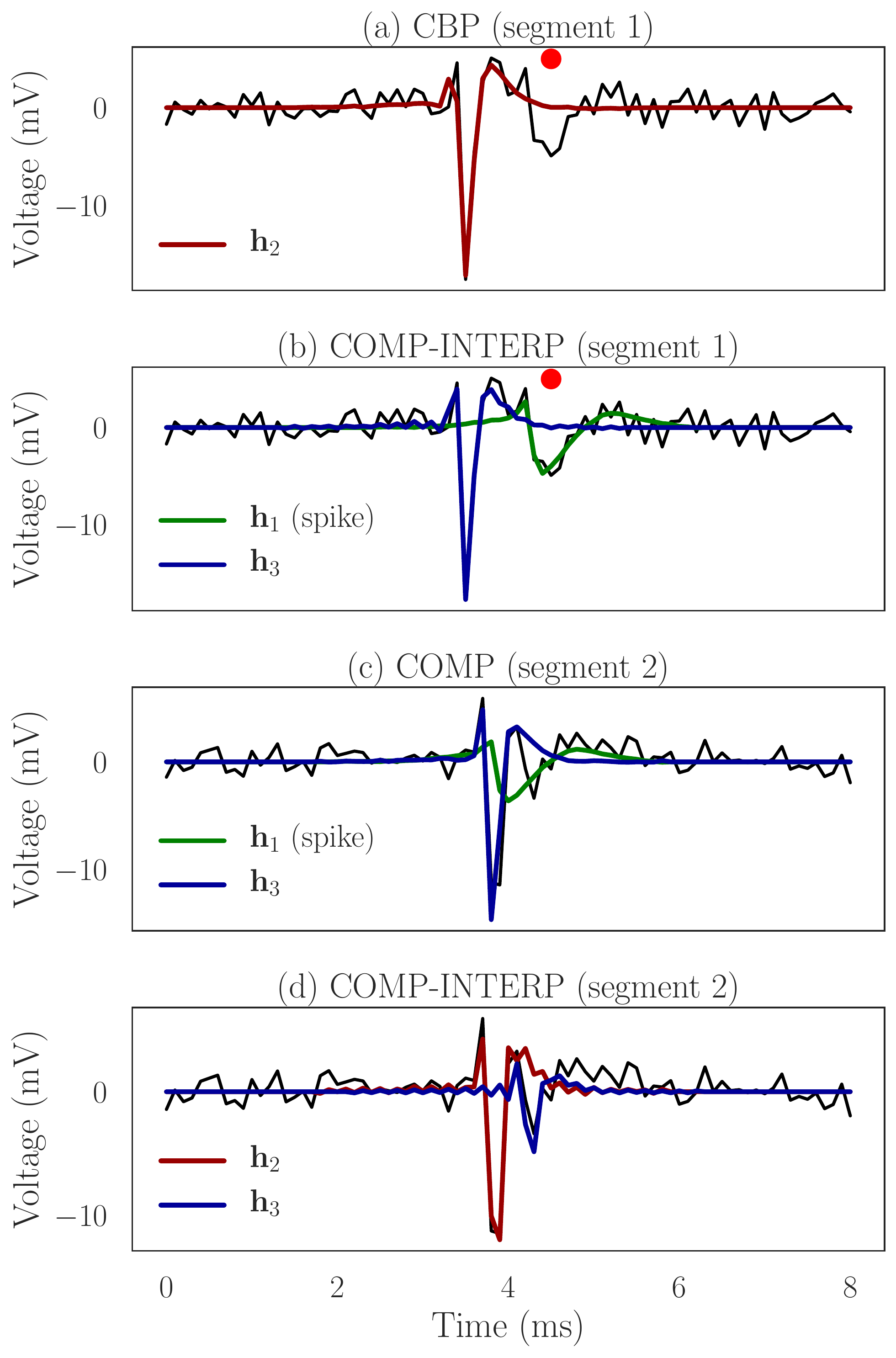}
	\caption{Example of applying CBP, COMP, and COMP-INTERP to segments of real data (black trace). The red dot shows where the true spike event occurs. The red, green and blue traces are reconstructions of the segments using only $\mathbf{h}_1$ (true spike template), $\mathbf{h}_2$  and $\mathbf{h}_3$, respectively. (a) A segment where CBP fails to correctly identify the occurrence of an event from the extracellular data, and (b) COMP-INTERP does. (c) A different segment where COMP incorrectly uses $\mathbf{h}_1$, but (d) COMP-INTERP does not.}
	\label{fig:real_CBP_COMP}
\end{figure}

Fig.~\ref{fig:real_error_curve} shows the result of sorting spikes associated with $\mathbf{h}_1$ using CBP, COMP and COMP-INTERP. The figure shows that the greedy approaches rival CBP, and are better in the low true miss rate regime. The true miss rate for CBP does not decrease below 5\%, even with thresholds of low amplitude. This indicates that CBP is not able to identify a subset of true events that COMP and COMP-INTERP identify correctly.\\ 
\noindent\textbf{Difference between $\ell_0$ and $\ell_1$}
The discrepancy in the true miss rate of CBP and COMP/COMP-INTERP motivated us to further examine segments for which the number of errors from CBP and the greedy methods differ. Fig.~\ref{fig:real_CBP_COMP} shows examples of such segments. 
Fig.~\ref{fig:real_CBP_COMP}(a) shows that CBP fails to capture the true spike event (red dot), resulting in a true miss event, whereas COMP-INTERP (Fig.~\ref{fig:real_CBP_COMP}(b)) uses $\mathbf{h}_1$ to correctly identify the event. The failure of CBP and the success of COMP-INTERP point to an important difference between the $\ell_1$ and $\ell_0$-based methods for the CSC step. To minimize its objective function, $\ell_1$-based CBP must strike a balance between the reconstruction error, which it can reduce by using additional templates, and the $\ell_1$ penalty, which can be reduced by using fewer templates or ones with lower amplitude. In Fig.~\ref{fig:real_CBP_COMP}(a), the choice of regularization parameter $\lambda$ is such that CBP chooses to use one template with large amplitude, thereby missing the true event.  Although a smaller regularization parameter $\lambda$ can be used to reduce the effect of the $\ell_1$ penalty, thereby identifying the true event and lowering the true miss rate, this would result in spurious events that would increase the number of false alarms. This discussion points to a limitation of $\ell_1$-based methods, namely the need to tune  $\lambda$ carefully. COMP-INTERP (and COMP), on the other hand, can select as many events as needed to make the reconstruction error below the estimate of the variance of the background noise. In Fig.~\ref{fig:real_CBP_COMP}(b), COMP-INTERP first selects $\mathbf{h}_3$, and then $\mathbf{h}_1$, which corresponds to the true event.

\noindent \textbf{Example of a non-integer shift} 
Fig. \ref{fig:real_CBP_COMP}(c) is an example of a segment where COMP raises a false alarm event, but COMP-INTERP does not (Fig. \ref{fig:real_CBP_COMP}(d)). COMP is forced to use the true spike template, $\mathbf{h}_1$, whereas COMP-INTERP uses $\mathbf{h}_3$ to select the secondary peak. This likely happens because the spike event does not occur on the discrete-time sampling grid, and further highlights the benefits of using the interpolated dictionary. That being said, we observe from Fig. \ref{fig:real_error_curve} that the two versions of COMP have similar performance, with  COMP-INTERP slightly outperforming COMP in the low true miss regime.

\begin{figure}
	\centering
	\includegraphics[width=\linewidth]{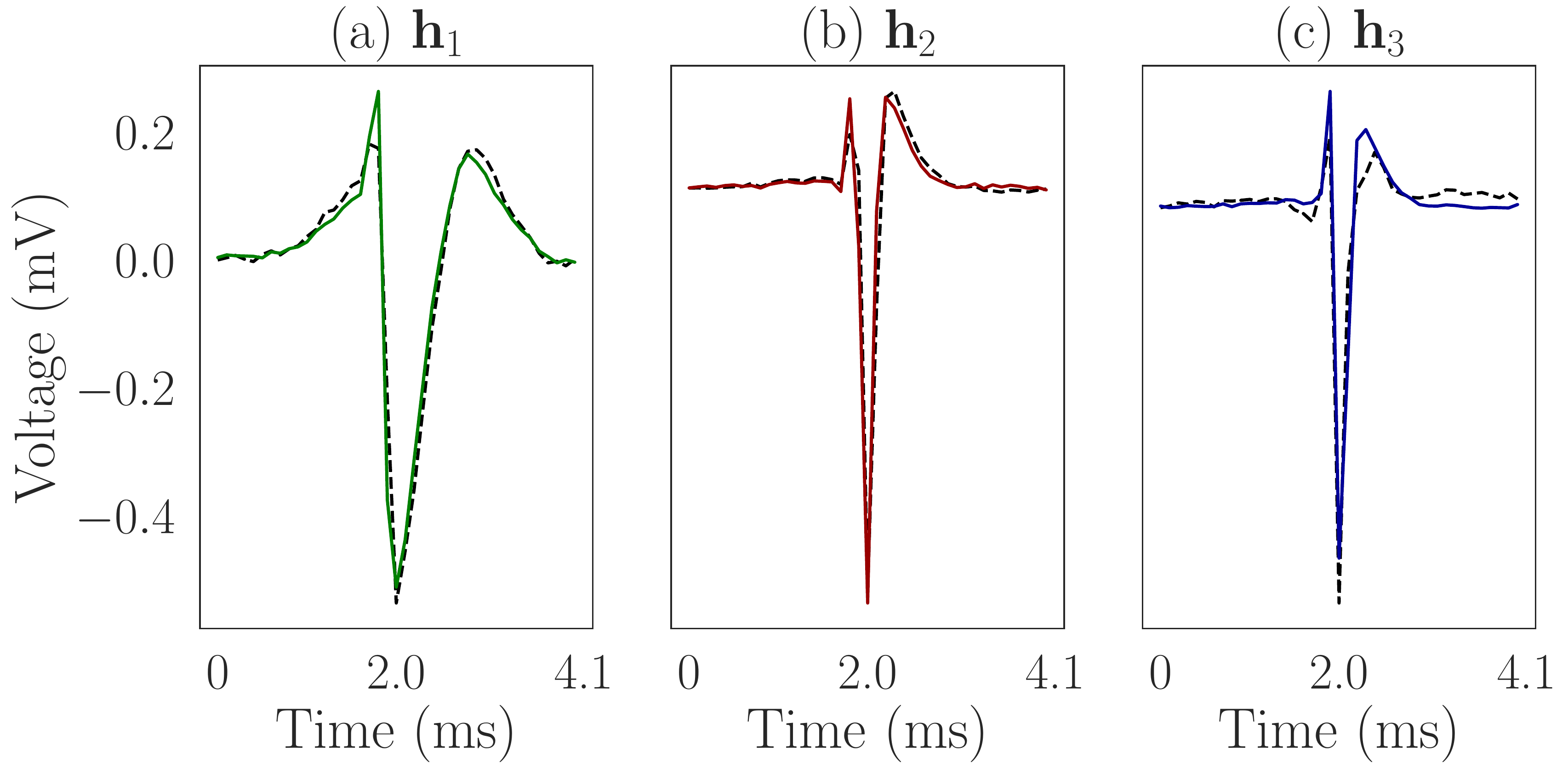}
	\caption{Initial $\mathbf{h}_c^{(0)}$ (black dashed trace), and learned templates $\mathbf{h}_c^{(\ast)}$ (colored trace). The values of the metric $\text{err}(\mathbf{h}_c^{(0)}, \mathbf{h}_c^{(\ast)})$ for $c=1,2,3$ were 0.21, 0.29, and 0.32, respectively.}
	\label{fig:real_templates}
\end{figure}

\noindent \textbf{Learned templates} Fig.~\ref{fig:real_templates} shows the templates that were learned by CDL using the extracellular data, and COMP-INTERP in the CSC step. The fact that the shapes of the learned templates are not significantly different from those of the initial templates suggests that we initialized the templates well. To determine how different the learned templates are from the initial ones, we computed $\text{err}(\mathbf{h}_c^{(0)}, \mathbf{h}_c^{(\ast)})$ for $c=1,2,3$, and then took the maximum of the three values. We found that the maximum equaled 0.32, which indicates that, although not obvious visually, the CDL algorithm did learn new templates.

\section{Discussion}\label{section:conclusion}
We have introduced novel Convolutional Sparse Coding (CSC) and Convolutional Dictionary Update (CDU) algorithms for a class of continuous-time signals that consist of time-shifted copies from multiple sources, each with its own template. The algorithms operate in discrete-time and construct an expanded, overcomplete, dictionary that accounts for the fact that the templates from the continuous-time model do not necessarily occur on the sampling grid of the discrete-time model. Specifically, the elements of the expanded dictionary consist of the original templates, along with their non-integer shifts, obtained by bandlimited interpolation.


The use of an expanded dictionary increases the computational demands of the algorithms, the CSC step in particular. To mitigate this, we chose to focus on greedy pursuit methods and proposed an efficient implementation of convolutional OMP (COMP) and COMP with interpolated dictionary (COMP-INTERP). The efficient COMP algorithm exploits the locality of the templates and is much faster than CBP, which is the state-of-the-art algorithm for CSC in continuous-time. Compared to convolutional MP, the overhead from the efficient COMP algorithm is negligible. We also demonstrated a novel algorithm to perform dictionary update with an expanded convolutional dictionary that accounts for non-integer delays. We showed empirically that more accurate templates were learned with the interpolated dictionary and that the results are robust across a wide range of SNRs. We conclude that our approach is a simple yet efficient paradigm for convolutional dictionary learning, that faithfully accounts for the continuous-time nature of the signal of interest.


We believe that the current work can be extended in several promising directions. First, we plan to extend this framework to the multivariate case. This will be useful for performing dictionary learning using recordings from sensor arrays, a setting in which the spatial information becomes as important as the temporal information of the signal. Then, we would like to mitigate COMP sparse coding errors that are due to template overlaps. Several interesting heuristics, such as bounded local optimization \cite{BLOOMP}, have been proposed for this.  These can be incorporated into our framework in efficient way. 
Next, since the CSC step is embarrassingly parallelizable across windows, we plan to implement the algorithm on GPU. 

\ifCLASSOPTIONcaptionsoff
  \newpage
\fi



\bibliographystyle{IEEEtran}


\bibliography{spikesorting}
\end{document}